\documentclass[article,twocolumn,floats,nofootinbib,nobibnotes,superscriptaddress,10pt]{revtex4-1}

\usepackage{graphicx,amsmath,amssymb,hyperref}
\hypersetup{colorlinks=true, linkcolor=red, citecolor=magenta, urlcolor=blue}
\usepackage{dcolumn}
\usepackage{multirow}
\usepackage{xcolor,colortbl}
\usepackage{enumitem}
\usepackage{eqnarray}
\usepackage{makecell}
\usepackage{cancel}
\usepackage{etoolbox}
\usepackage{tabularx}

\renewcommand{\arraystretch}{1.5}

\begin{document}
\title{Upcoming searches for decaying dark matter with ULTRASAT ultraviolet maps }

\author{Sarah Libanore}%
\email{libanore@bgu.ac.il}
\affiliation{Department of Physics, Ben-Gurion University of the Negev, Be'er Sheva 84105, Israel}

\author{Ely D. Kovetz}
\email{kovetz@bgu.ac.il}
\affiliation{Department of Physics, Ben-Gurion University of the Negev, Be'er Sheva 84105, Israel}

\begin{abstract}

Decaying dark matter (DDM)  can be tested via different astrophysical and cosmological probes. In particular, particles in the $\sim$\,9.5\,-\,30\,eV mass range that decay into monochromatic photons, would contribute to the extragalactic background light (EBL) in the ultraviolet (UV) bandwidth. 
In this work, we show that an intriguing improvement to the constraints on such DDM models can come from broadband UV surveys, such as GALEX or the upcoming ULTRASAT satellite. These provide diffuse light maps of the UV-EBL, integrated over a wide redshift range. The cross correlation between intensity fluctuations in these maps with a reference spectroscopic galaxy survey, can be used to reconstruct the redshift evolution of the EBL intensity; in this way, it is also possible to detect signatures of  contributions from DDM. We forecast the constraining power of (GALEX+ULTRASAT)$\times$DESI, and we show they will be able to detect DDM with decay rate up to $\mathcal{O}(10^{-26}\,{\rm s})$. In the context of axion-like particles (ALP), our forecasts can be converted to constraints on the ALP-photon coupling; our results show this technique will test ALP with coupling {$\lesssim\mathcal{O}(10^{-12}\,{\rm GeV^{-1}})$}, more than an order of magnitude better than current bounds in this mass range.

\end{abstract}

\maketitle

\section{Introduction}

The nature of dark matter (DM) is one of the biggest unknowns in modern cosmology. Reasonable explanations span an extremely wide scale of energies and include, among  others, a variety of solutions in which DM is ultimately made up of particles. These may only weakly interact with the standard model (see e.g.,~Refs.~\cite{Bertone:2004pz,Profumo:2017hqp} and references within), and can be detected either directly or indirectly, see e.g.~Refs.~\cite{Schumann:2019eaa,Gaskins:2016cha}.
Inside this zoo of possibilities, there is the chance that at least part of the DM is made up of particles which are capable to decay: these could be e.g.,~axions and axion-like particles~\cite{Preskill:1982cy,Abbott:1982af,Dine:1982ah,Turner:1989vc,Peccei:2008,Adams:2022pbo}, or sterile neutrinos~\cite{Hansen:2003yj,Kusenko:2009up,Dodelson:1993je}. Irrespective of their origin and their mass, broadly speaking we can refer to this class of models with the acronym DDM (decaying dark matter).

Constraints on DDM can be obtained from cosmological and astrophysical observations~\cite{Cadamuro:2011fd}; among  these, an important role is played by measurements of the extragalactic background light (EBL), namely the integrated emission of all the unresolved sources in the observed field. {\it In primis}, DDM models cannot predict energy injections that overcome the EBL amplitude: this provides an upper bound on their decay rate and abundance. A review of EBL measurements in the different frequency bands can be found e.g.,~in Refs.~\cite{Overduin:2004sz,Cooray:2016jrk,Hill:2018trh}. 

As for the ultraviolet (UV) region, EBL constraints still have large uncertainties, due to the
presence of foregrounds, e.g.,~zodiacal light, near-Earth airglow, Galactic dust~\cite{Hamden:2013,Murthy:2013,Henry:2014jga,Akshaya:2018}. Moreover, photons in this waveband get easily absorbed in the intergalactic medium, since their energy is close to the neutral-hydrogen ionization level. Measurements so far have been performed using the integrated light and photometric measurements from the Voyager 1 and 2 missions~\cite{Murthy:1999,Eldstein:2000}, Hubble~\cite{Brown:2000uq,Bernstein:2001sq} and the Galaxy Evolution Explorer (GALEX,~\cite{Martin:2004yr,Morrissey:2007hv}). 

Detection of excess in the UV band due to DDM is hence difficult; to get a clear indication that the emission has an extragalactic, cosmological origin, therefore, other techniques have to be used. In order to increase the chances for DDM detection, the optimal choice is to refer to surveys capable of collecting the integrated light from all bright and faint sources; this opportunity has grown in the recent years thanks to the rise of line intensity mapping science (LIM, see review in Refs.~\cite{Kovetz:2017agg,Bernal:2022jap} and DDM-related analyses in,~e.g.,~Refs.~\cite{Bernal:2022xyi,Bernal:2022wsu,Bernal:2022xyi,Creque-Sarbinowski:2018ebl,Liu:2019bbm,Facchinetti:2023slb}). 
In this context, for example, Ref.~\cite{Bernal:2020lkd} (\hyperlink{B20}{\color{magenta} B20} from here on) discussed how to detect DM decays from intensity maps treating them as interloper contributions. Interlopers can also provide information on other types of decaying particles, e.g.,~neutrinos with enhanced magnetic moment due to beyond the standard model physics~\cite{Bernal:2021ylz}.

The authors of Ref.~\cite{Creque-Sarbinowski:2018ebl}, instead, suggested to use cross-correlation to distinguish DM decays from the astrophysical EBL. In fact, when observing a wide cosmological volume, the EBL integrates emissions from galaxies that are 
located at different redshifts and emit in a broad frequency spectrum, hence EBL photons appear to be uniformly distributed in the volume. The correlation with a galaxy catalog, therefore, gets higher in the presence of DDM emissions in a certain frequency range, which trace the underlying DM field in a specific redshift shell. This condition, however, breaks if one can perform tomography of the EBL, namely reconstructing the redshift evolution of its intensity. In this way, the EBL intensity becomes itself a tracer of the DM distribution; to do so, clustering-based redshift can be used. 

The clustering-based redshift (CBR) technique, proposed originally in Refs.~\cite{Newman:2008mb,McQuinn:2013ib,Menard:2013aaa}, can be used to reconstruct the redshift evolution of the EBL intensity, by exploiting the cross correlation between a broadband survey and a reference spectroscopic galaxy survey. 
This kind of analysis relies on the comparison between the clustering properties of two biased tracers of the same underlying DM field; in our case, following Ref.~\cite{Libanore:2024wvv} (\hyperlink{LK24}{\color{magenta} LK24}), based on previous work by Ref.~\cite{Chiang:2018miw}\,(\hyperlink{C19}{\color{magenta} C19}), we consider as tracers the intensity fluctuations over a UV diffuse light map, and galaxies. 
The former are accessed by broadband surveys such as GALEX~\cite{Martin:2004yr,Morrissey:2007hv} and ULTRASAT~\cite{Sagiv:2013rma,ULTRASAT:2022,Shvartzvald:2023ofi}, while for the latter we rely on the forecasted DESI~\cite{DESI:2013agm,DESI:2016fyo,DESI:2016igz} 5\,years capability, which have been presented in Ref.~\cite{DESI:2023dwi}. 
The more galaxies and intensity fluctuations cluster, the more the CBR signal is high, since it is easier to link them to the same DM structure (i.e., the host DM halo), identifying the redshift the UV emission is sourced from. 

In this work, we build upon the analysis of \hyperlink{LK24}{\color{magenta} LK24}, with the goal of forecasting the capability of CBR in constraining the properties of DDM, when applied to UV broadband surveys. We always refer to the process in which the DDM particle decays into two monoenergetic photons, namely DDM$\to\gamma+\gamma$. The energy of the photons is set by the mass of the decaying particle; in our case, the bandwidth considered allows us to focus on DDM in the range $\sim 9.5$\,-\,$30\,{\rm eV}$. Extending the analysis to other processes, which involve daughter particles, is straightforward and can be done e.g.,~following the same approach in \hyperlink{B20}{\color{magenta} B20}. 
The main quantity of interest in our analysis is the decay rate of the particles, i.e.~the inverse of their lifetime; our forecasted constraints, however, need to account for degeneracies between this quantity and the branching ratio of the process, as well as with the abundance of DDM particles with respect to  all of the DM. 

If we specify the production mechanism, the decay rate can be related with the DDM-photon coupling constant; this can be done, e.g.,~in the case of axion-like particles (ALP, see, e.g.,~Refs.~\cite{Cadamuro:2011fd,Marsh:2015xka,Balazs:2022tjl,Adams:2022pbo} and therein).
ALPs are expected to be produced during spontaneous symmetry breaking, can have different coupling strengths, and may span a wide mass spectrum~\cite{Arias:2012az}; because of their characteristics, they represent an exciting DM candidate. 

The paper is structured as follows. In Sec.~\ref{sec:EBL} we review the formalism to model the UV-EBL: initially, we follow the approach in \hyperlink{C19}{\color{magenta} C19} and \hyperlink{LK24}{\color{magenta} LK24}, and then we introduce the contribution of decaying dark matter. With respect to \hyperlink{LK24}{\color{magenta} LK24}, we improve the modeling of the continuum escape fraction, relying on results from Refs.~\cite{Hayes:2010cj,Begley:2023}. In Sec.~\ref{sec:intensity} and Sec.~\ref{sec:CBR} we model our observables: we construct the intensity fluctuations in the maps of the broadband UV surveys, and we define its cross correlation with the galaxy survey, in order to perform the clustering-based redshift analysis. Sec.~\ref{sec:analysis} collects our forecasts for the constraining power on DDM; we compare our results with forecasts in a similar mass range obtained in \hyperlink{B20}{\color{magenta} B20} for LIM surveys. Finally, in Sec.~\ref{sec:ALP} we explicitly refer to ALPs, and we extend our constraints to their coupling strength. Following the model suggested in Ref.~\cite{Gong:2015hke}, we also consider ALPs with an extended mass spectrum. Overall, our forecasts show that diffuse light maps from GALEX and ULTRASAT and galaxy catalogs from DESI, will enable to reconstruct the redshift-dependent UV intensity. This kind of analysis will be capable of detecting emissions from DDM with masses $\sim 10$\,-\,$25\,{\rm eV}$, providing competitive constraints in a regime that is still widely unknown.

\section{Extragalactic Background Light}\label{sec:EBL}

\vspace*{-.1cm}
\subsection{Emissivity from astrophysical sources}
\vspace*{-.1cm}

The study of the extragalactic background light (EBL) in the near-ultraviolet (NUV) to far-ultraviolet (FUV) $\nu_{\rm obs} \sim 1$\,-\,$3\cdot 10^6\,{\rm GHz}$ ($\lambda_{\rm obs}\sim 3000$\,-\,$1000\,$\AA) bandwidth is challenging, mainly because of Galactic foreground emissions and efficient absorption in the intergalactic medium (IGM).
An interesting perspective has been suggested and applied by the authors of~\hyperlink{C19}{\color{magenta} C19}: the cross correlation between the  broadband, diffuse light map collected by GALEX and a reference spectroscopic galaxy survey, such as the Sloan Digital Sky Survey (SDSS,~\cite{SDSS:2004dnq,
Reid:2015gra}), can be used to reconstruct the redshift evolution of some observed integrated quantity. This is known as the clustering-based redshift technique~\cite{Newman:2008mb,McQuinn:2013ib,Menard:2013aaa}; in \hyperlink{C19}{\color{magenta} C19}, it was adopted to reconstruct the redshift evolution of the spatially averaged comoving volume emissivity, $\epsilon_{\rm EBL}(\nu,z)$ $[\rm erg\,s^{-1}Hz^{-1}Mpc^{-3}]$, as we summarize in Sec.~\ref{sec:CBR}.

To perform the analysis, \hyperlink{C19}{\color{magenta} C19} suggested to model the UV emissivity using the phenomenological expression
\begin{equation}
\label{eq:epsilon_ebl}
    \frac{\epsilon_{\rm EBL}(\nu,z)}{\epsilon_{1500}} = 
    \begin{cases}\smallskip
        &\left[\dfrac{\nu_{1216}}{\nu_{1500}}\right]^{\alpha_{1500}} \\\smallskip
        &\left[\dfrac{\nu_{1216}^{\alpha_{1500}-\alpha_{1100}}\nu^{\alpha_{1100}}}{\nu_{1500}^{\alpha_{1500}}}\right]  +\dfrac{{\mathcal{P}_{1216}\rm EW}}{\nu_{1216}\delta}\dfrac{\nu^2}{c} \\
        & \left[\dfrac{\nu_{1216}^{\alpha_{1500}-\alpha_{1100}}\nu_{912}^{\alpha_{1100}}\nu^{\alpha_{900}}}{\nu_{1500}^{\alpha_{1500}}\nu_{912}^{\alpha_{900}}}\right]f_{\rm LyC},
\end{cases}
\end{equation}
\vspace*{-.1cm}
where the $\nu$ subscripts indicate the rest frame wavelengths in \AA\, units, based on which frequencies are computed.\footnote{$\{\nu_{912},\nu_{1100},\nu_{1216},\nu_{1500}\}=\{3.29,2.72,2.46,2.00\}\times 10^6\,{\rm GHz}$.} We indicate with $\mathcal{P}_{1216}$ a piecewise function that is different from zero only when $|\nu - \nu_{1216}| <\nu_{1216}\delta$, with $\delta=0.005$. Eq.~\eqref{eq:epsilon_ebl} models three behaviours of the UV emissivity: the first row, which holds when the emitted rest frame frequency is smaller than the Lyman-$\alpha$ (Ly$\alpha$) line, $\nu < \nu_{1216}$, 
describes the non-ionizing continuum. In proximity of the line, 
$\nu_{1216}<\nu<\nu_{912}$, the non-ionizing continuum behaves as in the second row; on top of that, the Ly$\alpha$ line is parameterized by the equivalent width $\rm EW$. Finally, the third row describes the ionizing radiation above the Lyman break, $\nu> \nu_{912}$, where 
\begin{equation}\label{eq:fesc}
 \log_{10}f_{\rm LyC } = C_{\rm LyC}\log_{10}\left(\frac{1+z}{1+z_{\rm C 1}}\right) + \log_{10}f_{\rm LyC}^{z = z_{\rm C 1}}
\end{equation}
accounts for the escape fraction of ionizing photons from galaxies, and it determines the presence of the Lyman break in the UV spectrum. We define 
\begin{equation}\label{eq:C_lyC}
   C_{\rm LyC} = \frac{\log_{10}f_{\rm LyC}^{z = z_{\rm C 2}}-\log_{10}f_{\rm LyC}^{z = z_{\rm C 1}}}{\log_{10}[(1+z_{\rm C 2})/(1+z_{\rm C 1})]}\,,
\end{equation}
to scale $f_{\rm LyC}$ with respect to its pivot values at $z_{\rm C1},z_{\rm C2}$. 

\begin{figure}[ht!]
    \centering
    \includegraphics[width=\columnwidth]{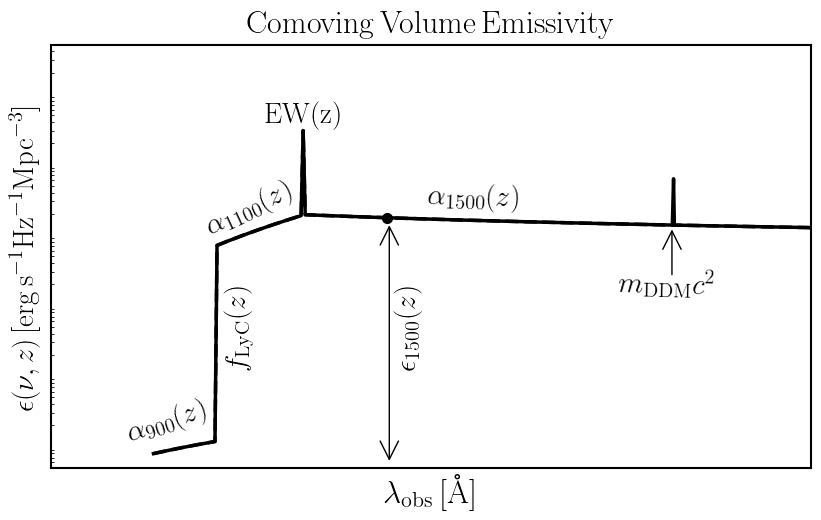}
    \caption{Summary of the model for the rest frame emissivity UV-EBL adopted in this work, when DDM is described by a single mass. The continuum and Ly$\alpha$ line are modeled through Eq.~\eqref{eq:epsilon_ebl}, while the DDM line position and intensity depend on $m_{\rm DM}c^2$, as described in Eqs.~\eqref{eq:nu_dm} and~\eqref{eq:eps_DM}. }
    \label{fig:emissivity}
\end{figure}

\noindent
In Eq.~\eqref{eq:epsilon_ebl}, the emissivity is normalized by $\epsilon_{1500} = \epsilon_{1500}^{z=0}(1+z)^{\gamma_{1500}}$; redshift and frequency dependencies enter here, in the Ly$\alpha$ equivalent width 
\begin{equation}
{\rm EW} = C_{\rm Ly\alpha}\log_{10}\left[({1+z})/({1+z_{\rm EW 1}})\right] + {\rm EW}^{z = z_{\rm EW 1}},
\end{equation}
where $C_{\rm Ly\alpha}$ is defined analogously to Eq.~\eqref{eq:C_lyC}, and in the exponents 
$\alpha_{1500} = \alpha_{1500}^{z=0}+C_{1500}\log_{10}(1+z)$, $\alpha_{1100} = \alpha_{1100}^{z=0}+C_{1100}\log_{10}(1+z)$, $\alpha_{900} = \alpha_{900}^{z=0}$. 

{The functional form of Eq.~\eqref{eq:epsilon_ebl} resembles the spectral energy distribution of a typical galaxy, see e.g.,~the outputs of the BEAGLE simulation~\cite{Chevallard_2016}. Its amplitude, instead, should be estimated by integrating the contributions of all galaxies. A possible alternative to model the UV-EBL would be to rely on the halo model: once the UV emissivity is provided for each halo, it can be used to scale the halo mass function; the clustering of the EBL, in this case, would be described by a bias factor, obtained by scaling the halo bias. While on the semi-analytical level both our method and the halo model require approximations and assumptions, the halo model approach has the advantage of being suitable to work with simulations; we will investigate possibility this in future work. }
Tab.~\ref{tab:fid} summarizes the fiducial values of the parameters adopted in the analysis in Sec.~\ref{sec:analysis}; these are set to the posterior results of \hyperlink{C19}{\color{magenta} C19}, apart from $f_{\rm LyC}^{z_{\rm C1},z_{\rm C2}}$. 
\hyperlink{C19}{\color{magenta} C19} found upper limits for the average value of this parameter at $z_{\rm C1,C2}=\{1,2\}$, while measurements on single galaxies at lower redshift have been collected e.g.,~in Refs.~\cite{Flury:2022asy,Flury:2022jsy}. Constraining the average evolution of $f_{\rm LyC}(z)$ is quite challenging, but a good probe can be found in the correlation between this parameter and the escape fraction of Ly$\alpha$ photons, $f_{\rm Ly\alpha}(z)$. This has recently been done in Ref.~\cite{Begley:2023}, which provided the relation $f_{\rm LyC}\simeq 0.15\times f_{\rm Ly\alpha}$; in the following, we rely on this and on the $f_{\rm Ly\alpha}$ measurements at $z= \{0.3,2.2\}$ in Refs.~\cite{Cowie:2009mt,Hayes:2010zn}, collected by Ref.~\cite{Hayes:2010cj}. These $f_{\rm LyC}(z)$ values are well below the limit established in \hyperlink{C19}{\color{magenta} C19}, smaller than the ones used in \hyperlink{LK24}{\color{magenta} LK24}.

\begin{table}[ht!]
    \centering
\renewcommand{\arraystretch}{1.2}
    \begin{tabular}{|cc||cc|}
    \hline
     Parameter & Fiducial value &  Parameter & Fiducial value \\
    \hline
    $\log_{10}\epsilon_{1500}^{z=0}$  &  25.63 & 
    $\gamma_{1500}$ & 2.06  \\
    $\alpha_{1500}^{z=0}$ & -0.08 &
    $C_{1500}$ & 1.85  \\
    $\alpha_{1100}^{z=0}$ & -3.71 &
    $C_{1100}$ & 0.50 \\
    $\alpha_{900}^{z=0}$ & -1.5 & 
    ${\rm EW}^{z=0.3}$ & -6.17\,\AA \\
    ${\rm EW}^{z=1}$ & 88.02\,\AA &
    ${\rm EW}^{z=2}$ & 176.7\,\AA  \\
    $z_{\rm C1}$ & 0.3 & $z_{\rm C2}$ & 2.2\\
    $\log_{10}f_{\rm LyC}^{z=0.3}$ & -3.4 &
    $\log_{10}f_{\rm LyC}^{z=2.2}$ & -2.1  \\
    \hline
    $\gamma_{\nu}$ & -0.86&
    $\gamma_z$ & 0.79 \\
    $b_{1500}^{z=0}$ & 0.32 & & \\
    \hline
    \end{tabular}
    \smallskip 
    \caption{Fiducial values of the parameters of the rest frame EBL emissivity $\epsilon_{\rm EBL}(\nu,z)$ in Eq.~\eqref{eq:epsilon_ebl} and its bias $b_{\rm EBL}(\nu,z)$ in Eq.~\eqref{eq:bias_J}. Values from Ref.~\cite{Chiang:2018miw}; $f_{\rm LyC}^{z_{\rm C1,C2}}$ from Refs.~\cite{Cowie:2009mt,Hayes:2010zn}.}
    \label{tab:fid}
\end{table}

Figs.~\ref{fig:emissivity} and~\ref{fig:comp_em} show how the rest frame emissivity looks like for fixed values of $z$ and $\nu$, respectively.  Overall, the prescription in Eq.~\eqref{eq:epsilon_ebl} is in good agreement with the model in Ref.~\cite{Haardt:2011xv}, which accounts for galaxy emission and dust extinction; for this reason, we consider it as the astrophysical component of the EBL.

As a final remark, we note that the model we adopted relies on the average comoving emissivity, which does not discriminate between properties of different astrophysical UV sources. A more refined analysis can be realized relying on the halo model formalism, as e.g.,~Refs.~\cite{Lidz:2011dx,Breysse:2014uia}. In such approach, the emissivity is modeled as a function of the properties of the source galaxies and their host DM halos. We will develop this model in a follow-up work; for the aims of this paper, instead, we take advantage of the agnostic point of view that Eq.~\eqref{eq:epsilon_ebl} provides, with respect to the choice of the galaxy population. The analysis  in \hyperlink{C19}{\color{magenta} C19} proved the reliability of this approach.

\subsection{Decaying Dark Matter Emissivity}\label{sec:emiss_DDM}

\begin{figure}[ht!]
    \centering
    \includegraphics[width=\columnwidth]{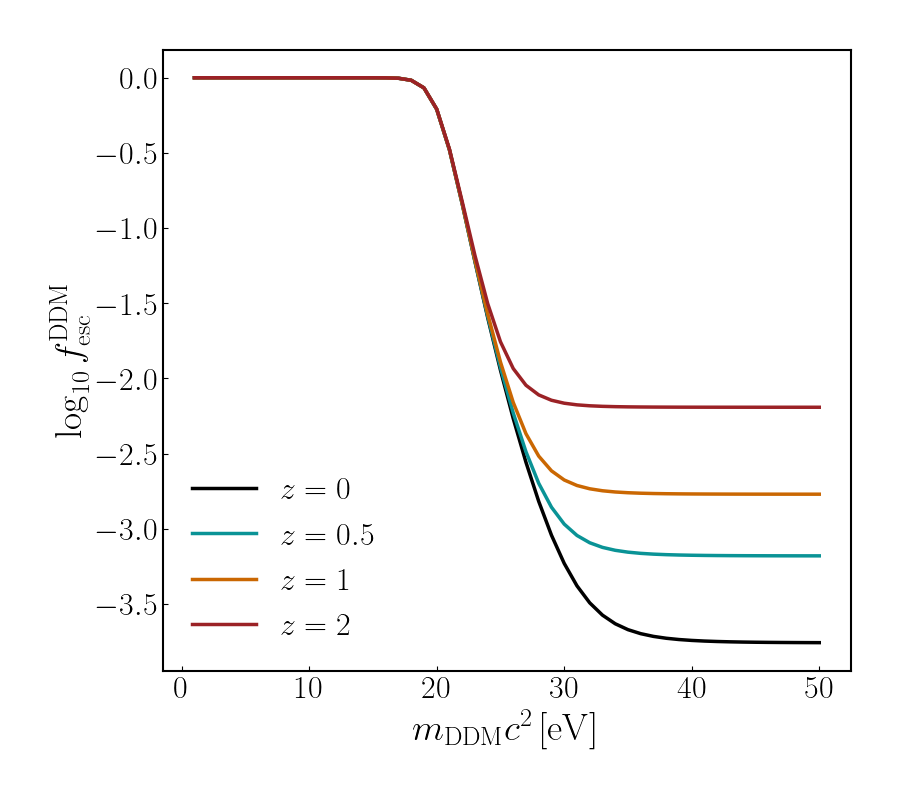}
    \caption{$\log_{10}f^{\rm DDM}_{\rm esc}$ for different $z$ and DDM masses; the value at high masses reaches $f_{\rm LyC}(z)$. The parameter enters the computation of the DDM emissivity in Eq.~\eqref{eq:eps_DM}.}
    \label{fig:DM_esc}
\end{figure}

While Eq.~\eqref{eq:epsilon_ebl} parameterizes the UV-EBL from astrophysical sources, extra contribution may come from exotic sources, such as decaying dark matter particles (DDM). 
These can constitute either part or the totality of the DM content of the Universe; we parameterize its abundance through the ratio between the DDM and total DM density parameters,
\begin{equation}\label{eq:fddm}
    f_{\rm DDM} = \Omega_{\rm DDM}/\Omega_{c}.
\end{equation} 
If at least part of the DM consists of decaying particles, these produce extra radiation that contributes to the EBL in a specific frequency band; throughout this work, we always refer to the DDM$\to\gamma+\gamma$ process, in which a single DDM particle decays into two monoenergetic photons.
Thence,  radiation produced in the decay of a particle with mass $m_{\rm DDM}c^2$ has rest-frame frequency  
\begin{equation}\label{eq:nu_dm}
  \nu_{\rm DDM} = \frac{m_{\rm DDM}c^2}{4\pi\hbar}\,,
\end{equation}
where $\hbar$ is the Planck constant divided by $2\pi$.
Generalizations to the case in which DDM decays e.g.,~to a photon and a daughter particle can be straightforwardly derived by rescaling $\nu_{\rm DDM}$ by the ratio of the masses (see e.g.,~discussion in \hyperlink{B20}{\color{magenta} B20}).
If DDM has mass $\mathcal{O}(10\,{\rm eV})$, the frequency of the photons it produces falls in the UV bandwidth; in this case, DDM contributes to the UV-EBL emissivity with an emission line, standing out above the non-ionizing continuum as we show in Fig.~\ref{fig:emissivity}.

To estimate the contribution of DDM to the UV-EBL and hence to the observed intensity, we model the DDM specific comoving-volume emissivity following Refs.~\cite{Chen:2003gz,Finkbeiner:2011dx,Pierpaoli:2003rz,Creque-Sarbinowski:2018ebl,Bernal:2020lkd}. For a fixed DDM mass, we define
\begin{equation}\label{eq:eps_DM}
    \epsilon_{\rm DDM}(m_{\rm DDM},z)= \frac{\Theta_{\rm DDM}}{\nu_{\rm DDM}}f^{\rm DDM}_{\rm esc}\left[ \Omega_{c}\rho_c \right]e^{-\tau_{\rm DDM}^{-1}(t-t_0)}\,,
\end{equation}
where we neglected stimulated decays (which would anyway increase the emission rate,\footnote{{Ref.~\cite{Caputo:2018vmy}.shows that accounting for stimulated decay introduces a $(1+2\mathcal{F}_\gamma)$ factor in Eq.~\eqref{eq:eps_DM}, where $\mathcal{F}_\gamma$ is the photon occupation number of the radiation field that stimulates the emission. Photon emission in the UV range may be stimulated by the UV-EBL or by resolved UV-bright sources, which are masked out from our all-sky maps. The contribution from the UV-EBL would enhance the DDM signal while also introducing further correlation between the cosmological and astrophysical components; a detailed modeling of this effect is beyond the scope of this paper.}} so we remain conservative), and we introduced $\rho_c=3H_0^2/(8\pi G)$ as the critical density and $f^{\rm DDM}_{\rm esc}$ as the escape fraction of the photons produced in the decay. Finally, the quantity
\begin{equation}\label{eq:theta_DDM}
    \Theta_{\rm DDM} = f_{\gamma\gamma}f_{\rm DDM}\tau_{\rm DDM}^{-1}\,,
\end{equation}
encloses the information about the process: $f_{\rm DDM}$ is the fraction defined in Eq.~\eqref{eq:fddm}, $f_{\gamma\gamma}=1$ is the branching ratio of the $\rm DDM\to \gamma+\gamma$ process, and $\tau_{\rm DDM}^{-1}$ its decay rate, namely the inverse of the particle lifetime. We collect these parameter together since they are degenerate in the analysis in Sec.~\ref{sec:analysis}. 
 To avoid DM abundance evolving relevantly across cosmic time, we require $\tau_{\rm DDM}^{-1} < 10^{-18}\,{\rm s}^{-1} = H_0^{-1}$, where $H_0^{-1}$ is the age of the Universe. Since $e^{-\tau_{\rm DDM}^{-1}(t-t_0)}\sim 1$, the only redshift evolution in $\epsilon_{\rm DDM}(m_{\rm DDM},z)$ is hence due to $f^{\rm DDM}_{\rm esc}$. 
We parameterize this factor as  
\begin{equation}\label{eq:fesc_DDM}
\begin{aligned}
    f^{\rm DDM}_{\rm esc}(\nu_{\rm DDM},z) =& f_{\rm LyC}(z)+(1-f_{\rm LyC}(z))\,\times\\
    &\times \mathcal{S}\left[\frac{\log_{10}(\lambda_{\rm DDM}/1216\,{\rm \AA})}{50\,{\rm dex}}\right],
\end{aligned}
\end{equation}
where $f_{\rm LyC}(z)$ is the parameter adopted in the astrophysical case, Eq.~\eqref{eq:fesc}, while $\mathcal{S}[x]= (1+e^{-x})^{-1}$ is a sigmoid varying smoothly between 0 and 1, centered around the Ly$\alpha$ rest-frame wavelength. The reason behind this modeling is to constrain $f_{\rm esc}^{\rm DDM}$ to the same values as $f_{\rm LyC}(z)$, since photons are absorbed similarly irrespective if they originated from galaxies or DDM. The $f_{\rm LyC}$ factor, however, does not account for $\nu$ dependence: in the astrophysical case, this is characterized by the presence of the Lyman break in the last line of Eq.~\eqref{eq:epsilon_ebl}; in the DDM case, instead, we take care of it through the sigmoid in Eq.~\eqref{eq:fesc_DDM}.

\begin{figure}[ht!]
    \centering
    \includegraphics[width=\columnwidth]{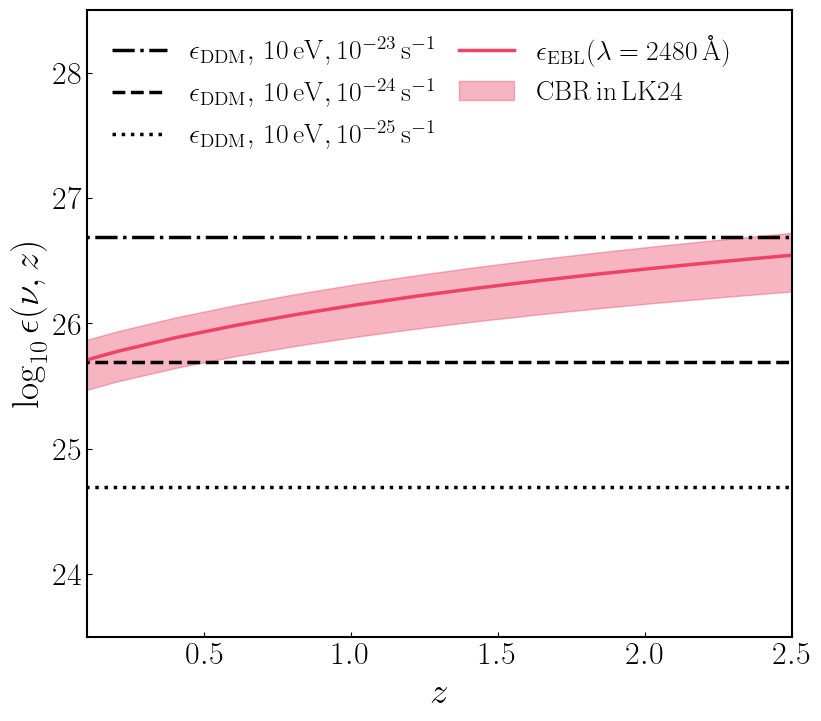}
    \caption{$\epsilon_{\rm DDM}$ compared with $\epsilon_{\rm EBL}$ as a function of $z$, for a fixed rest frame frequency. Here we consider $m_{\rm DDM}c^2=10\,{\rm eV}$; larger masses would decay into photons with higher frequency and produce smaller emissivity (e.g.,~when $m_{\rm DDM}c^2=20\,{\rm eV}$, the emitted rest frame frequency is $\lambda = 1240\,{\rm \AA}$ and the dot-dashed line ($10^{-23}\,{\rm s}^{-1}$) gets lower than $ \epsilon_{\rm EBL}$ at $z\sim 1$). We show the $1\sigma$ error estimated in Ref.~\cite{Libanore:2024wvv} for (GALEX+ULTRASAT)$\times$DESI with conservative bias prior. Note that the $\lambda$ at which the emissivity is  computed here, differs from the one shown in the discussion section of Ref.~\cite{Libanore:2024wvv}.}
    \label{fig:comp_em}
\end{figure}

Fig.~\ref{fig:DM_esc} shows $f^{\rm DDM}_{\rm esc}$ for different DDM masses: above $m_{\rm DDM}c^2 \sim 20\,{\rm eV}$, photons produced in the decays are easily absorbed. 
In addition to this local contribution, we need to account for further absorption due to intervening hydrogen clouds along the line-of-sight; their effect is modeled in Eq.~\eqref{eq:dJnu} through the $e^{-\zeta(\nu,z)}$ factor. 

The approach described so far is different from, e.g.,~\hyperlink{B20}{\color{magenta} B20}, which limits the analysis at $m_{\rm DDM}c^2 < 20\,{\rm eV}$ and collects the contributions of $f_{\rm esc}^{\rm DDM}$ and $e^{-\zeta(\nu,z)}$ inside a parameter $\tilde{\Theta}_{\rm DDM}$. Analogously to ${\Theta}_{\rm DDM}$ in Eq.~\eqref{eq:theta_DDM}, this also accounts for $f_{\gamma\gamma}$, $f_{\rm DDM}$ and $\tau_{\rm DDM}^{-1}$. In order to compare our results with theirs, in the second part of the analysis in Sec.~\ref{sec:analysis_compareLIM} we thus fix $f_{\rm esc}^{\rm DDM}(\nu_{\rm DDM},z) \equiv 1$,  $e^{-\zeta(\nu,z)} \equiv 1$, and we rely on $\tilde{\Theta}_{\rm DDM}$ instead of ${\Theta}_{\rm DDM}$.

As an illustrative example, in Fig.~\ref{fig:comp_em} we compare the astrophysical UV-EBL $\epsilon_{\rm EBL}$ with the DDM emissivity $\epsilon_{\rm DDM}$ in the rest-frame frequency emitted by a $m_{\rm DDM}c^2=10\,{\rm eV}$ particle. The two contributions are comparable when $\Theta_{\rm DDM} \simeq 5\times 10^{-24}\,{\rm s^{-1}}$; constraints on the astrophysical UV-EBL in the figure are obtained following the procedure in \hyperlink{LK14}{\color{magenta} LK14}. 

The rest-frame emissivity depicted in Fig.~\ref{fig:comp_em} is not the main observable we rely on. In fact, the broadband surveys we analyse in this work, collect the contribution of $\epsilon_{\rm EBL}$ and $\epsilon_{\rm DDM}$ once they get redshifted over a broad range of frequencies. Therefore, the quantity they observe is integrated over different $\nu_{\rm obs}$, as described in Sec.~\ref{sec:intensity}: in this case, each part of the UV spectrum, including the DDM line, can be captured in the diffuse light once they are emitted from different redshifts, and with a different intensity. Overall, the presence of DDM increases the observed flux in the detector with respect to the only-EBL case. By reconstructing its redshift evolution (see Sec.~\ref{sec:CBR}), it is possible to capture signatures of the DDM also in the $z$-shape of the reconstructed quantity. All these effects and the signatures that the presence of DDM produces on the observable of the broadband survey, are described in the next section.

\section{Observed EBL intensity in broadband survey}\label{sec:intensity}

As described in Sec.~\ref{sec:EBL}, the DDM emissivity in Eq.~\eqref{eq:eps_DM} represents a further contribution to the UV-EBL, which sums to emissions with astrophysical origin, in
Eq.~\eqref{eq:epsilon_ebl}. The total UV comoving-volume emissivity hence is
\begin{equation}\label{eq:epsilon}
    \epsilon(\nu,z)= \epsilon_{\rm EBL}(\nu,z)+\epsilon_{\rm DDM}(\nu,z).
\end{equation}
This quantity can be used to estimate the observed specific intensity $J_{\nu_{\rm obs}}$ [Jy/sr]. Using the radiative transfer function in an expanding Universe~\cite{Gnedin:1996qr} and introducing the detector response function $R(\nu_{\rm obs})$,\footnote{The response function is defined in terms of the normalized quantum efficiency {$\mathcal{Q}_E(\lambda_{\rm obs})$}, i.e.,~the number of electrons produced by the detector per incident photon depending on its frequency. In our analysis in Sec.~\ref{sec:analysis}, we retrieve the GALEX response functions from the public repository \url{http://svo2.cab.inta-csic.es/svo/theory/fps3/}, while we compute the ULTRASAT response function {as $R(\lambda_{\rm obs})= \mathcal{Q}_E N_S A_S$, where $N_S=4$ is the number of sensors in the detector focal plane and $A_S=(4.5\,{\rm cm})^2$ their effective area. We retrieved these information from the ULTRASAT official website and Refs.~\cite{Bastian_Querner_2021,Asif:2021vrm}. The response function is then renormalized to get $\int_{\lambda_{\rm obs}^{\rm min}}^{\lambda_{\rm obs}^{\rm max}}d\lambda_{\rm obs}{R(\lambda_{\rm obs})}/{\lambda_{\rm obs}} = 1$.}} the redshift-dependent specific intensity becomes
\begin{equation}\label{eq:dJnu}
\begin{aligned}
    \frac{dJ_{\nu_{\rm obs}}}{dz}=&\frac{c}{4\pi H(z)(1+z)}\,\,\times\\
    &\times \int_{\nu_{\rm obs}^{\rm min}}^{\nu_{\rm obs}^{\rm max}}\frac{d\nu_{\rm obs}}{\nu_{\rm obs}}R(\nu_{\rm obs})\epsilon(\nu,z)e^{-\zeta(\nu,z)}\,,
\end{aligned}
\end{equation}
where $c$ is the speed of light, $H(z)$ the Hubble factor at the redshift of interest $z$, and $\nu_{\rm obs} = \nu/(1+z)$ the observed frequency at the detector. The $e^{-\zeta(\nu,z)}$ factor describes absorptions the photons can undergo while traveling along the line-of-sight. The effective optical depth $\zeta(\nu,z)$ has been modeled in Refs.~\cite{Madau:1995js,Inoue:2014zna}, based on prescriptions for the line-of-sight probability distribution of hydrogen clouds and their column density.

A high-quality measurement of ${dJ_{\nu_{\rm obs}}}/{dz}$ would provide us insightful information on the integrated sources of the UV-EBL and their evolution across cosmic time; however, such measurement is not straightforward. 
In this work, following \hyperlink{C19}{\color{magenta} C19} and \hyperlink{LK24}{\color{magenta} LK24}, we focus on the capability of broadband surveys, such as GALEX~\cite{Martin:2004yr,Morrissey:2007hv} and ULTRASAT~\cite{Sagiv:2013rma,ULTRASAT:2022,Shvartzvald:2023ofi}, in constraining this quantity. Ref.~\cite{Murthy:2014} builds a diffuse UV-light map based on data from the All Sky and Medium Imaging Surveys performed by GALEX, masking resolved sources to estimate the combined contribution of the UV-EBL and foregrounds; \hyperlink{LK24}{\color{magenta} LK24} suggested that a similar procedure can be applied to the almost-full sky map ULTRASAT will realize in the first six months of the mission, as well as to the map realized based on the data of the low-cadence survey. Ref.~\cite{Scott:2021zue} suggested a similar approach using forecasts for the Cosmological Advanced Survey Telescope for Optical and UV Research (CASTOR,~\cite{Cote:2019}).

In the study of the diffuse light map obtained from a broadband survey, the quantity that is indeed observed is the specific intensity integrated over the full observed frequency range, namely
\begin{equation}\label{eq:intensity}
    {J}_{\nu_{\rm obs}} = \frac{c}{4\pi}\int_{\nu_{\rm obs}^{\rm min}}^{\nu_{\rm obs}^{\rm max}} \frac{d\nu_{\rm obs}{R}(\nu_{\rm obs})}{\nu_{\rm obs}}\int_0^\infty \frac{dz\,\epsilon(\nu,z)e^{-\zeta(\nu,z)}}{H(z)(1+z)}\,,
\end{equation}
together with its spatial fluctuations across different directions on the sky.
While on one side, such observable prevents knowing the redshift evolution of the UV-EBL, on the other side it is extremely powerful, since it collects light emitted from all its sources, including the faintest ones, potentially providing a huge amount of astrophysical and cosmological information. 

A possible way to extract its redshift evolution, as proposed in \hyperlink{C19}{\color{magenta} C19}, is through the CBR technique, as described in Sec.~\ref{sec:CBR}. To apply it, we need to model the clustering of the fluctuations in the intensity broadband map; this can be done using the effective intensity-weighted bias 
\begin{equation}\label{eq:bias_J}
\begin{aligned}
b_J(z) &= \biggl[\int_{\nu_{\rm obs}^{\rm min}}^{\nu_{\rm obs}^{\rm max}} \frac{d\nu_{\rm obs}}{\nu_{\rm obs}}{R(\nu_{\rm obs})}\,e^{-\zeta(\nu,z)}\times \\
&\times [b_{\rm EBL}(\nu,z)\epsilon_{\rm EBL}+b_{\rm DDM}(\nu,z)\epsilon_{\rm DDM}]\biggr]\times\\
& \times \biggl[\int_{\nu_{\rm obs}^{\rm min}}^{\nu_{\rm obs}^{\rm max}} \frac{d\nu_{\rm obs}}{\nu_{\rm obs}}{R(\nu_{\rm obs})}e^{-\zeta(\nu,z)}[\epsilon_{\rm EBL}+\epsilon_{\rm DDM}]\biggr]^{-1},
\end{aligned}
\end{equation}
where we omitted the $(\nu,z)$ dependencies in the emissivity for brevity. In the previous equation, we defined 
\begin{equation}
    b_{\rm EBL}(\nu,z) = b_{1500}^{z=0}\left[{\nu}/{\nu_{1500}}\right]^{\gamma_{b_\nu}}(1+z)^{\gamma_{b_z}}.
\end{equation}
as the bias of the astrophysical sources, with fiducial values $\{b_{1500}^{z=0},\gamma_{b_\nu},\gamma_{b_z}\} = \{0.32,-0.86,0.79\}$ as described in Tab.~\ref{tab:fid} (see more detail in \hyperlink{C19}{\color{magenta} C19} and \hyperlink{LK24}{\color{magenta} LK24}). For the DDM part, we set $b_{\rm DDM} \equiv 1$, since decays should perfectly trace the underlying DM field. 

The use of a broadband survey offers an interesting advantage in the study of DDM: the wide range of frequencies observed allows us to collect photons produced in decays that take place at different $z$, increasing the amount of signal collected.  Since the DDM emissivity has a different redshift dependence than the astrophysical EBL, reconstructing the $z$-dependent intensity $dJ_{\nu_{\rm obs}}/dz$ would allow us to increase the sensitivity of the instrument to $\Theta_{\rm DDM}$ cases that yield $\epsilon_{\rm DDM}(\nu_{\rm DDM},z)<\epsilon_{\rm EBL}(\nu,z)$. 

\begin{figure}[ht!]
    \centering
    \includegraphics[width=\columnwidth]{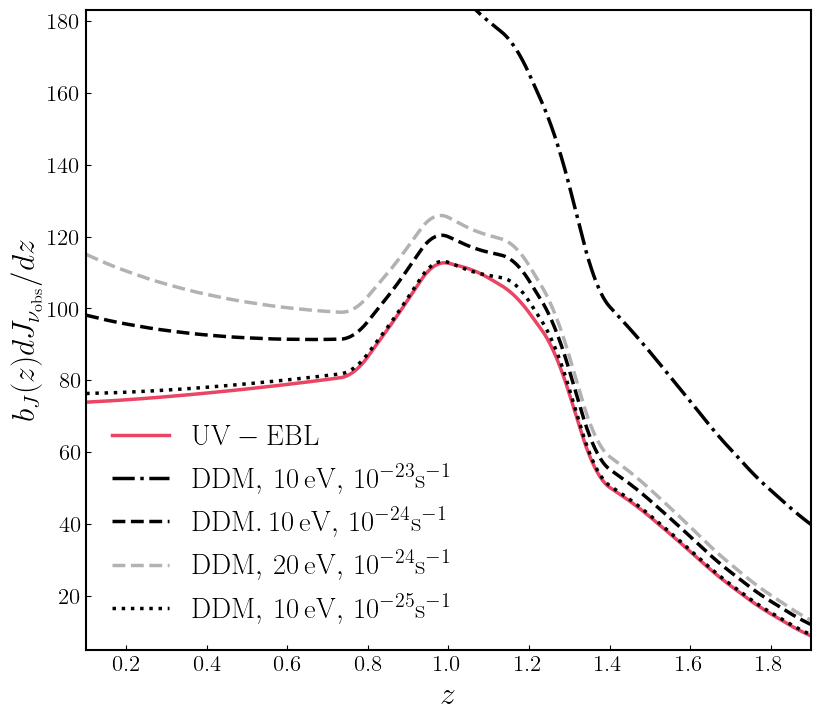}
    \caption{$b_J(z)dJ_{\nu_{\rm obs}}/dz$ computed using Eqs.~\eqref{eq:dJnu} and~\eqref{eq:bias_J}; this is the quantity the CBR technique reconstructs (see Sec.~\ref{sec:CBR}), and it represents the redshift evolution of the UV flux, weighted by the response function of the detector and integrated over its broadband frequency range. Here we show the ULTRASAT case; the pink, continuous line is modeled from the UV-EBL with astrophysical origin, while other lines show signatures due to DDM with different $m_{\rm DDM}c^2$, $\Theta_{\rm DDM}$. {The amplitude and shape of the UV-EBL is affected by uncertainties in the astrophysical parameters; the interested reader can compare with Fig.~3 in Ref.~\cite{Chiang:2018miw} to understand how each parameter affects the final result.}}
    \label{fig:bdJ}
\end{figure}

To do so, a possible solution is to use the clustering based redshift technique, as detailed in the next section. This technique is capable of reconstructing the redshift evolution of $b_J(z)dJ_{\nu_{\rm obs}}/dz$; prescriptions or measurements on the bias of the UV sources can then be used to disentangle the two quantities. We show in Fig.~\ref{fig:bdJ} how the quantity $b_J(z)dJ_{\nu_{\rm obs}}/dz$ looks like for ULTRASAT. Here, we compare the case in which only the UV-EBL from astrophysical sources is accounted, with results obtained adding DDM contributions. For dependencies on the other parameters and for other detectors, see Fig.~3 in \hyperlink{C19}{\color{magenta} C19} and  Fig.~1 in \hyperlink{LK24}{\color{magenta} LK24}.

\section{Clustering-Based Redshift}\label{sec:CBR}

The redshift evolution of $dJ_{\nu_{\rm obs}}/dz$ in Eq.~\eqref{eq:dJnu} and of the EBL volume emissivity in Eq.~\eqref{eq:epsilon} can be reconstructed using the clustering based redshift (CBR,~\cite{Newman:2008mb,McQuinn:2013ib,Menard:2013aaa}) technique. 
This is based on the study of the cross-correlation across different directions in the sky between the $J_{\nu_{\rm obs}}$ integrated intensity (Eq.~\eqref{eq:intensity}) observed by a UV broadband survey and the number density of a reference spectroscopic galaxy survey. The amplitude of the correlation allows us to split in redshift bins the UV intensity and to reconstruct its redshift evolution. At the same time, the cross-correlation isolates the signal from extragalactic contributions, while foreground emissions, being uncorrelated with the galaxy survey, only contribute to the noise, as we describe in Sec.~\ref{sec:noise}.

Following the approach detailed in \hyperlink{C19}{\color{magenta} C19} and \hyperlink{LK24}{\color{magenta} LK24}, we model the CBR observable as a two-point correlation function smoothed over a certain angular distance $\theta$. This can be expressed as  
\begin{equation}\label{eq:wJg_z}
 \bar{\omega}_{\tilde{J}{\rm g}}(z) = b_J(z)b_g(z)\frac{dJ_{\nu_{\rm obs}}}{dz}\int_{\theta_{\rm min}(z)}^{\theta_{\rm max}(z)}d\theta\,W(\theta) \omega_{m}(\theta,z).
\end{equation}
The observable is modulated by the intensity of the EBL, $dJ_{\nu_{\rm obs}}/dz$, and by the clustering properties of the two tracers, encapsulated in the bias parameters: $b_J(z)$ relies on Eq.~\eqref{eq:bias_J}, while for $b_g(z)$ we refer to the different forecasted catalogs described in Ref.~\cite{DESI:2023dwi}. 
A further contribution is then provided by the angular two-point function of the underlying DM field~\cite{Maller:2003eg},
\begin{equation}
    \omega_{m}(\theta,z) = \frac{1}{2\pi}\int_0^\infty dk\,kP_m(k,z)\mathcal{J}_0(k\theta\chi(z))\frac{dz}{d\chi},
\end{equation}
where $\mathcal{J}_0$ is the Bessel function of the first type, $\chi(z)$ the radial comoving distance and $dz/d\chi = H(z)/c$. $P_m(k,z)$ is the DM power spectrum; we compute it using the public library \texttt{CAMB}\footnote{\url{https://github.com/cmbant/CAMB}}~\cite{Challinor:2011bk}, using the halofit model prescription from Ref.~\cite{Mead:2020vgs}. 
Finally, the normalized window function $W(\theta)\propto \theta^{-0.8}$ marginalizes over angular distances, between the minimum $\theta_{\rm min}$ defined to avoid the too-strong non-linear clustering regime, and $\theta_{\rm max}$ which is set in order to have a stable calibration over the field. We set $\theta_{\rm min}(z) = \arctan[r_{\rm min}/D_A(z)]$, where $D_A(z)$ is the cosmological angular diameter distance. The value of the physical distance $r_{\rm min} = 0.5\,{\rm Mpc}$ is chosen in \hyperlink{C19}{\color{magenta} C19} to avoid too strong non-linear clustering on small scales. Further considerations on this and on how results change using different values for $r_{\rm min}$, can be found in \hyperlink{LK24}{\color{magenta} LK24}. The choice of $\theta_{\rm max}$ instead depends on the detector; values used in our analysis are collected in Tab.~\ref{tab:specs}.

Following the analysis in \hyperlink{LK24}{\color{magenta} LK24}, in this work we forecast the constraining power of the combination of GALEX and ULTRASAT diffuse light maps in cross correlation with DESI 5\,yr results as presented in Ref.~\cite{DESI:2023dwi}. 
Table~\ref{tab:specs} collects the relevant information of the surveys; note that GALEX provides two filters, in the near-UV (NUV) and far-UV (FUV) bands, while ULTRASAT will focus its analysis  only on the NUV band, reaching at least $\sim 10$\,times better sensitivity than GALEX~\cite{Shvartzvald:2023ofi}. On the other hand, ULTRASAT, whose launch is expected in 2026, will provide two diffuse light maps useful for this kind of analysis: a reference {\it full sky map}, built in the first six months of the mission, and the map associated with the {\it low cadence survey}. The latter will observe a smaller sky area, but it will be deeper: lower magnitudes will be reached thanks to longer, integrated observational times and repeated passages on the same field.  

\begin{table}[ht!]
    \centering
    \begin{tabular}{|c|cccc|}
    \hline
      &  $\lambda_{\rm obs}\,[{\rm \AA}]$ & $m_{\rm AB}$ &  $A_{\rm pix}$ & $\theta_{\rm max}\,$[deg]\\
    \hline
    GALEX FUV & [1350,\,1750] & 22 & $(50''/{\rm pix})^2$ & $[0.2,\,0.7]$\\
    GALEX NUV & [1750,\,,2800] & 22 & $(50''/{\rm pix})^2$ & $[0.2,\,0.7]$ \\
    {ULTRASAT} & {[2300,\,2900]} & 23.5 & {$(5.45\,''/{\rm pix})^2$} & {4} \\
        \hline
    \end{tabular}
    \begin{tabular}{|c|cccccc|}
    \hline
       &  $\Omega_{\rm survey} \,{\rm [deg^2]}$ & $N_g$ & $\lambda/\Delta \lambda$ & $\Delta z$ &  $z_{\rm min}$& $z_{\rm max}$ \\
         \hline 
        DESI & 14000 & $\mathcal{O}(10^7)$ & 1000 &  0.1 & 0.1 & 2\\ 
    \hline
    \end{tabular}
    \caption{Specs for the (almost) full-sky, broadband surveys (top, GALEX~\cite{Martin:2004yr} and ULTRASAT~\cite{Shvartzvald:2023ofi}): observed bandwidth, limiting magnitude, area of the effective pixels used in the analysis and maximum angular distance inside which we consider the calibration stable enough to perform the cross correlation. Specs for the galaxy survey (bottom, DESI~\cite{DESI:2023dwi}): observed sky area, total number of galaxies, spectroscopic resolution, width of the redshift bins and redshift range for the analysis. We set $z_{\rm max} = 2$ since broadband surveys can only detect UV emission up to this distance (see Fig.~\ref{fig:bdJ}).}
    \label{tab:specs}
\end{table}

\subsection{Noise computation}\label{sec:noise}

To evaluate the UV detector noise variance, as well as the noise due to CBR, we follow~\hyperlink{LK24}{\color{magenta} LK24}.
First of all, we estimate the noise variance of the EBL map produced by a survey with limiting AB magnitude $m_{\rm AB}$. This is converted into flux density per each pixel through
\begin{equation}
    f_{\nu}^{\rm pix} = \frac{8.90-m_{\rm AB}}{2.5\,A_{\rm pix}},
\end{equation}
where $A_{\rm pix}$ is the pixel surface area; grouping pixels into ``effective pixels" with larger area is a useful strategy to reduce $\sigma_N^2$. 
 Consequently, the noise variance per pixel is
\begin{equation}\label{eq:sigma_N}
    \sigma_N^2 = \left[{f_\nu^{\rm pix}}/{5}\right]^2 .
\end{equation}
{Eq.~\eqref{eq:sigma_N} is used as a proxy of the uncertainty in the measurement of the observed flux, while at this stage we neglect further sources of instrumental noise.}

With respect to the CBR technique described in Sec.~\ref{sec:CBR}, \hyperlink{LK24}{\color{magenta} LK24} estimates the noise by expanding the approach in Refs.~\cite{Newman:2008mb,Menard:2013aaa}, and computing the number of pixel-galaxy pairs that can be created inside the area $\pi\theta_{\rm max}^2$ and redshift bin in which CBR is applied. This is 
\begin{equation}\label{eq:pairs_cbr}
    N_{\rm pairs} = N_{\rm pix}N_{g,i}^{\theta_{\rm max}} = 
    \frac{N_{g,i}}{A_{\rm pix}}\frac{(\pi\theta_{\rm max}^2)^2}{\Omega_{\rm survey}}\,,
\end{equation}
where $N_{\rm pix}$ is the number of pixels and $N_{g,i}^{\theta_{\rm max}}$ the number of galaxies per redshift bin in that area. These numbers are computed knowing, respectively, the pixel area $A_{\rm pix}$, the observed number density $dN_g/dzd\Omega$ and the overlapping sky area $\Omega_{\rm survey}$ between the UV and galaxy survey footprints. 
The number of patches over which the analysis is performed is $N_\theta = \Omega_{\rm survey}/\pi\theta_{\rm max}^2$. To improve the signal-to-noise ratio in the analysis, overlapping patches and tithering could also be used. However, this would require a non-trivial modeling for the covariance; for the aim of the Fisher analysis we discuss in Sec.~\ref{sec:analysis}, we thus decided to rely only on non-overlapping pairs. 

In our baseline analysis we consider ULTRASAT and GALEX full-sky maps, therefore the $\Omega_{\rm survey}$ area is set by the galaxy survey, as described in Tab.~\ref{tab:specs}. As for ULTRASAT, a valuable alternative is to rely on the low-cadence survey instead: this reaches higher sensitivity ($m_{\rm AB} \sim 24.5$), while surveying a smaller sky patch ($\sim 6800\,{\rm deg}^2$). As described in \hyperlink{LK24}{\color{magenta} LK24}, in the context of CBR analysis, the full-sky and low-cadence survey maps provide comparable noise and forecast results.
In Sec.~\ref{sec:analysis} we only refer to the full-sky map, but results on the low-cadence survey can be derive straightforwardly. 

The CBR noise has to be weighted by the variance of the intensity measurements in the map $\sigma_J^2$: this is estimated combining the noise variance $\sigma_N^2$ in Eq.~\eqref{eq:sigma_N} with the variance of the intensity itself. We assume that the measurement in each pixel is drawn from a Poisson distribution with mean $J_{\nu}^{\rm obs} = J_{\nu_{\rm obs}}^{\rm DDM}+J_{\nu_{\rm obs}}^{\rm EBL}(1+\mathcal{A}_{\rm fg})$, where $J_{\nu_{\rm obs}}^{\rm DDM},J_{\nu_{\rm obs}}^{\rm EBL}$ are the monopoles of the DDM and UV-EBL emissivity respectively, computed as in Eq.~\eqref{eq:intensity}. The term $J_{\nu_{\rm obs}}^{\rm EBL}\mathcal{A}_{\rm fg}$ represents the contribution due to foregrounds: this has been introduced because near-Earth airglow, zodiacal light, Galactic dust, and other components introduce an offset between the expected UV-EBL monopole of astrophysical origin and the observed one~\cite{Akshaya:2018}. \hyperlink{LK24}{\color{magenta} LK24} estimated this parameter to be $\mathcal{A}_{\rm fg}=1.8$ for GALEX FUV and $2.2$ for GALEX NUV and ULTRASAT. Overall, the intensity variance in the map is  
$    \sigma_J^2 = [J_{\nu_{\rm obs}}^{\rm DDM}+J_{\nu_{\rm obs}}^{\rm EBL}(1+\mathcal{A}_{\rm fg})]^2 +\sigma_N^2$.

The CBR noise is finally estimated to be 
\begin{equation}\label{eq:noise}
\begin{aligned}
    \mathcal{N}_{\rm CBR} &= \frac{\delta z_i}{\delta z_c}\sqrt{\frac{\sigma_J^2}{N_{\rm pairs}N_{\theta}}} \\
    &= \frac{\delta z_i}{\delta z_c}\sqrt{\frac{A_{\rm pix}{\bigl[J_{\nu_{\rm obs}}^{\rm DDM}+J_{\nu_{\rm obs}}^{\rm EBL}(1+\mathcal{A}_{\rm fg}) + \sigma_N^2\bigr]}}{\pi\theta_{\rm max}^2N_{g,i}}},
\end{aligned}
\end{equation}
where $\delta z_i / \delta z_c$ compares the size of the spectroscopic bins of the survey with the typical scale of matter clustering. %, which we set to $\delta z_c\sim 10^{-2}$: bins larger than this quantity are less efficient in the estimate of the CBR redshift.
Ref.~\cite{Menard:2013aaa} suggests to choose $\delta z_i \sim \delta z_c\sim 10^{-3}$ to optimize the analysis; following this and considerations in \hyperlink{LK24}{\color{magenta} LK24}, we set $\delta z_c = {\rm max}[10^{-3},H(z)(1+z)r_{\rm min}]$ to also account for the redshift evolution of this scale ($r_{\rm min}$ is the same as in Sec.~\ref{sec:CBR}). We finally define the width of the spectroscopic bins $\delta z_i$ as the larger value between $\delta z_c$ and the instrumental resolution $\lambda/\Delta \lambda$ in Tab.~\ref{tab:specs}.

\section{Analysis}\label{sec:analysis}

In the following, we apply the CBR technique detailed in the previous section in the context of the Fisher forecast analysis presented in \hyperlink{LK24}{\color{magenta} LK24}. The goal is to understand up to which value of $\Theta_{\rm DDM}$ (see Eq.~\eqref{eq:theta_DDM}) we will be able to constrain DDM by using the correlation between (GALEX+ULTRASAT) diffuse UV-light maps and the DESI galaxy catalog. Cross correlation has been proven to be a powerful tool to constrain DDM in the different mass ranges, see e.g.,~Refs.~\cite{Creque-Sarbinowski:2018ebl,Bartlett:2022ztj,Paopiamsap:2023uuo}.

We here rely on ULTRASAT full-sky map, but the analysis can be extended to its low-cadence survey~\cite{Shvartzvald:2023ofi}. We start by considering generic DDM with a monochromatic mass distribution, while in Sec.~\ref{sec:ALP} we will investigate axion-like particles (ALPs), both with monochromatic and broad mass spectra.

\subsection{Fisher forecasts}

\begin{figure}[ht!]
\includegraphics[width=\columnwidth]{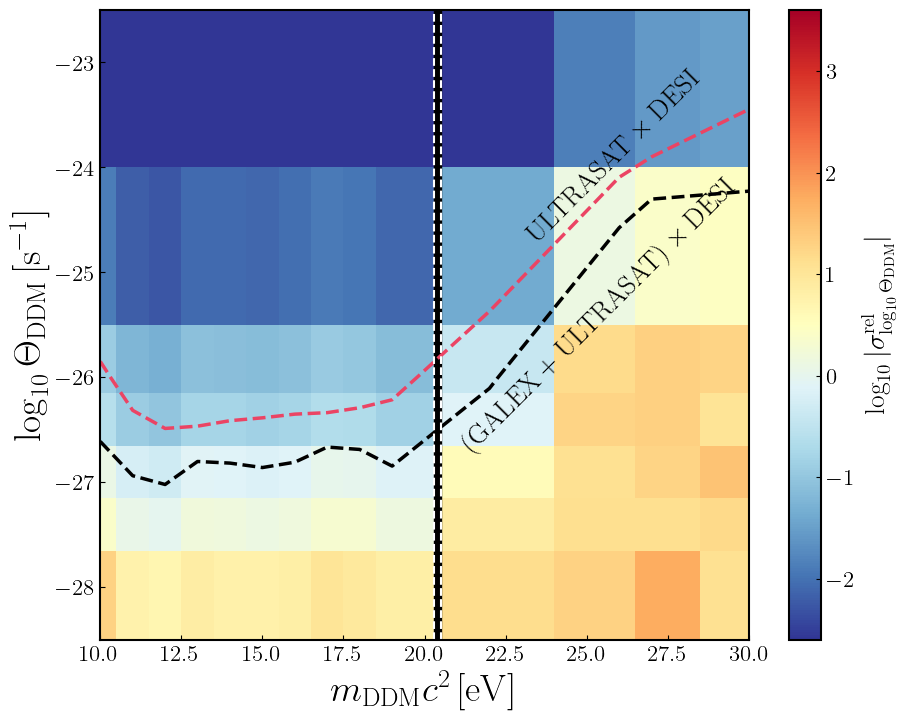}
    \caption{Forecasted relative errors on $\Theta_{\rm DDM}$ from our Fisher analysis for CBR applied to (GALEX+ULTRASAT)$\times$DESI, as a function of the particle mass $m_{\rm DDM}c^2$. Values above the black dashed contour line (blue region) can be detected at $2\sigma$; the magenta dashed line shows how the constraints worsen when only ULTRASAT data are used. The dashed vertical lines indicate particles with masses $2\nu_{1216}(1-\delta)\leq m_{\rm DDM}c^2 \leq 2\nu_{1216}(1-\delta)$, where $\delta$ is defined in Eq.~\eqref{eq:epsilon_ebl}: these would decay in photons having frequency in the proximity of the Ly$\alpha$ line, hence their detection is more challenging. }
    \label{fig:sigma_theta_DDM}
\end{figure}

To forecast the constraining power of the CBR technique, we rely on the Fisher matrix formalism~\cite{Vogeley:1996,Tegmark:1996bz},
\begin{equation}\label{eq:fisher}
    F_{\alpha\beta} = \sum_{i}\frac{\partial \bar{\omega}_{\tilde{J}g}(z_i)}{\partial\vartheta_\alpha}\frac{\partial \bar{\omega}_{\tilde{J}g}(z_i)}{\partial\vartheta_\beta}\frac{1}{\mathcal{N}_{\rm CBR}^2(z_i)},
\end{equation}
where we sum over the $i$-redshift bins of the CBR analysis for each detector-galaxy survey pair. The range and width of the $z$ bins are defined in Tab.~\ref{tab:specs}, while the derivatives of the signal are computed using Eq.~\eqref{eq:wJg_z}, and the noise $\mathcal{N}_{\rm CBR}(z_i)$ is estimated in Eq.~\eqref{eq:noise}. We compute the Fisher matrix separately for GALEX FUV$\times$DESI, GALEX NUV$\times$DESI and ULTRASAT$\times$DESI and we sum them to get the final results. This is done under the assumption that datasets from different detectors are uncorrelated; this is an approximation and in real data analysis the use of the full covariance will partially deteriorate the final results. For this reason, as a pessimistic forecast, we also quote results for ULTRASAT$\times$DESI alone. Constraints from the full analysis will fall between these and the (GALEX+ULTRASAT)$\times$DESI ones.

In our Fisher forecast, the parameter set we use is 
\begin{equation}\label{eq:vartheta}
\begin{aligned}
    \vartheta=&\{\log_{10}[b\epsilon]_{1500}^{z=0},\,\gamma_{1500}, \,\gamma_{\nu},\,\gamma_z,\\
    & \alpha_{1500}^{z=0},\,C_{1500},\,\alpha_{1100}^{z=0},\,C_{1100},\,\alpha_{900}^{z=0},\\
    &{\rm EW}^{z=z_{\rm EW1}},\,{\rm EW}^{z=z_{\rm EW2}},\\
    &\log_{10}f_{\rm LyC}^{z=z_{\rm C1}},\,\log_{10}f_{\rm LyC}^{z=z_{\rm C2}},\\
    &\log_{10}\Theta_{\rm DDM}\},
\end{aligned}
\end{equation}
which collects the parameters introduced in Sec.~\ref{sec:EBL} and the DDM parameter $\Theta_{\rm DDM}$. The DDM bias $b_{\rm DDM}$ and the galaxy biases $b_g(z)$ are held fixed to their fiducial values throughout the analysis. Note that, following \hyperlink{C19}{\color{magenta} C19} and \hyperlink{LK24}{\color{magenta} LK24}, we collected the local emissivity $\epsilon_{1500}^{z=0}$ and the bias amplitude $b_{1500}^{z=0}$ in a single parameter, $\log_{10}[b\epsilon]_{1500}^{z=0}$. The CBR analysis, in fact, is only capable of reconstructing the product between the two, through $b_J(z)dJ_{\nu_{\rm obs}}/dz$; the two parameters can then be separated by measuring the clustering bias independently, e.g.,~from resolved sources in the map (see \hyperlink{C19}{\color{magenta} C19} for detail). Based on results in \hyperlink{LK24}{\color{magenta} LK24}, we adopt conservative Gaussian priors on the EBL bias parameters $\{\sigma_{b_{1500}^{z=0}},\sigma_{\gamma_\nu},\sigma_{\gamma_z}\}$ = $\{0.05,1.30,0.30\}$; this allows us to break internal degeneracies between parameters. Moreover, we add a prior $\sigma_{\alpha_{900}} = 0.1$ on the shape of the ionizing continuum: due to the small escape fraction introduced in Sec.~\ref{sec:EBL}, the CBR can not constrain this parameter.
The marginalized error on each $\vartheta_\alpha$ is then estimated from the inverse of the Fisher matrix in Eq.~\eqref{eq:fisher}, as $\sigma_\alpha = \sqrt{F^{-1}_{\alpha\alpha}}$.  \vspace*{-.5cm}

\subsubsection{Results}

We run the analysis for DESI in cross correlation separately with GALEX NUV, GALEX FUV and ULTRASAT, and we finally sum the Fisher matrices to get the overall result. For each case, we consider DDM particles with masses $9.5\,{\rm eV}\leq m_{\rm DDM}c^2\leq 30\,{\rm eV}$, which are held fixed throughout each run. The upper bound is chosen according to the discussion on the escape fraction in Sec.~\ref{sec:emiss_DDM}; the lower bound instead is set by the shortest frequency observed by ULTRASAT, $\nu_{2900}$: since in our analysis $z_{\rm min} = 0.1$, we set $(m_{\rm DDM}c^2)_{\rm min}$ from Eq.~\eqref{eq:nu_dm}, using the rest frame frequency $\nu_{\rm DDM} = (1+z_{\rm min})\nu_{2900}$.

Fig.~\ref{fig:sigma_theta_DDM} shows our results: for different values of $\log_{10}\Theta_{\rm DDM}$ ranging from $-23$ to $-28$, we plot the relative marginalized errors {$\sigma^{\rm rel}_{\log_{10}\Theta_{\rm DDM}}=\sqrt{F^{-1}_{\log_{10}\Theta_{\rm DDM}\log_{10}\Theta_{\rm DDM}}}/\log_{10}\Theta_{\rm DDM}$} obtained based on our Fisher analysis; since we neglected stimulated decays, our forecasts reasonably show a conservative bound. 
The black contour line in Fig.~\ref{fig:sigma_theta_DDM} indicates values that can be detected at $2\sigma$ for (GALEX+ULTRASAT)$\times$DESI{, namely $\sigma^{\rm rel}_{\log_{10}\Theta_{\rm DDM}}=1/2$}; for masses below 20\,eV, these are $\log_{10}\Theta_{\rm DDM}\in [{-26},{-27}]$. These values are smaller than the value required to get $\epsilon_{\rm DDM}(\nu,z)> \epsilon_{\rm EBL}(\nu,z)$: this implies that this analysis will detect signatures of DDM even without a clear detection of excess in the UV-EBL; detecting decays close to the Ly$\alpha$ line will be more challenging. 
The magenta contour line in Fig.~\ref{fig:sigma_theta_DDM} shows how results deteriorate if we rely on ULTRASAT$\times$DESI; even if results lose almost an order of magnitude, the CBR still provides good constraints. 

\begin{figure}[ht!]
    \centering
    \includegraphics[width=\columnwidth]{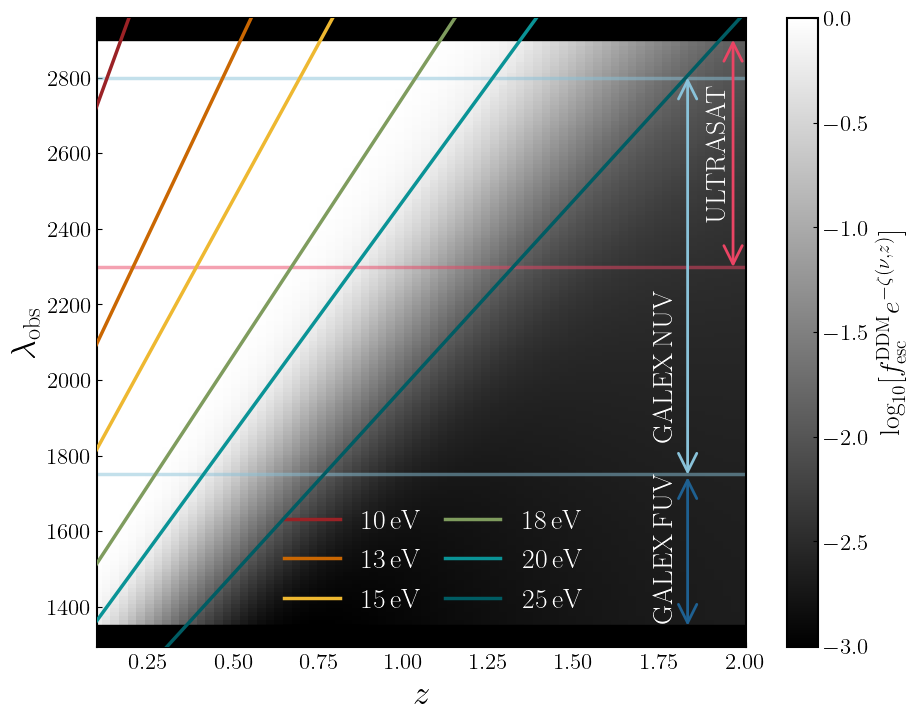}
    \caption{Contribution of different DDM decays to the broadband UV region observed by GALEX and ULTRASAT. Coloured lines indicate the redshifted wavelengths emitted by different masses; the underlying gray colour describes the absorptions due to $f_{\rm esc}^{\rm DDM}$ and $e^{-\zeta(\nu,z)}$ (see Sec.~\ref{sec:emiss_DDM}); darker color indicates more absorption, thus less signal. As we describe in the text, between $m_{\rm DDM}c^2\sim 9.5$\,-\,$12\,{\rm eV}$, the detected signal increases because decays from heavier particles can be detected from a wider redshift range; above $\sim 18\,{\rm eV}$, absorption become non-negligible, hence the signal decreases. }
    \label{fig:signal_evolution}
\end{figure}

It is interesting to note how the constraining power changes across the mass range. Constraints initially improve between $m_{\rm DDM}c^2 \sim\,$9.5\,eV and 12\,eV, then slightly worsen and stabilize between $13$\,-\,$18$\,${\rm eV}$, decreasing again afterwards.
This trend can be explained as an interplay between the $z$-evolution of the DDM signal, the properties of the bandwidth of the UV-detectors and the galaxy survey.
On one side, in fact, the photons produced by DDM particles with larger masses can be collected from a wider $z$ range, since their frequency enters sequentially the GALEX FUV, GALEX NUV and ULTRASAT bands. This would lead to a larger signal for higher masses; at the same time, however, the signal contribution lowers at higher $z$, since the flux decreases and because of the evolution in the values of the absorption $e^{-\zeta(\nu,z)}$. Fig.~\ref{fig:signal_evolution} can help visualizing these facts. 

The drop around $m_{\rm DDM}c^2= 13$\,eV also mirrors the shape of the galaxy number density. As described in detail in \hyperlink{LK24}{\color{magenta} LK24}, the DESI 5\,yr forecast in Ref.~\cite{DESI:2023dwi} accounts for different catalogs in the different redshift ranges: in particular, between $z\sim 0.1$\,-\,$0.4$ the number density is larger thanks to the presence of bright galaxies (BGS), while between $z\sim 0.4$\,-\,$1.1$ it lowers, being dominated by luminous red galaxies (LRG). Emissions from $m_{\rm DDM}c^2=10$\,eV particles in the BGS region can be observed by ULTRASAT, which has a higher response function at $\lambda_{\rm obs}\sim 2600\,{\rm \AA}-2800\,{\rm \AA}$ (see \hyperlink{LK24}{\color{magenta} LK24}). On the contrary, photons produced at low $z$ by $m_{\rm DDM}c^2=$13\,eV particles get captured by GALEX NUV, while they reach the more sensitive ULTRASAT band only when sourced from higher $z$, in the LRG region. Overall, this combines in a more noisy CBR measurement in the $m_{\rm DDM}c^2\sim$13\,eV case, and hence to a larger constraining power on lower masses, until the $z$ range from which photons are collected gets large enough to overcome this difference.

\subsubsection{Comparison with future line-intensity mapping surveys}\label{sec:analysis_compareLIM}

The observable we discussed in the previous sections is based on the analysis of intensity fluctuations in UV maps. It is reasonable, therefore, to compare its constraining power to forecasts for upcoming and future line intensity mapping (LIM) surveys, e.g.,~from \hyperlink{B20}{\color{magenta} B20}.

LIM surveys observe selected emission lines from particle interactions in the Universe; the choice of the observed frequency range sets the line that is observed and the redshift from which this is produced. LIM surveys can be used to trace the large scale structures of the Universe (see Refs.~\cite{Kovetz:2017agg,Bernal:2022jap} for review), using line emissions due to star formation inside galaxies (e.g.,~CO or CII...), or processes in the IGM (e.g.,~Ly$\alpha$, H$\alpha$...). The intensity measured in a line-intensity map is due to all the unresolved sources in the observed sky area, and it is usually analysed through summary statistics of its fluctuations, such as the one-point correlation function (also referred to as {voxel intensity distribution}, VID) and the two-point correlation function (i.e.,~the {power spectrum}, $P(k)$). 

If (part of) the DM decays in the local Universe, its emission may constitute an interloper for atomic or molecular lines that are sourced at higher redshift, thus introducing further anisotropies in the summary statistics built from their observations. As a consequence, the LIM VID and power spectrum will be affected by the photons produced in the decay. 
With respect to the mass range we are interested in this work, \hyperlink{B20}{\color{magenta} B20} shows that constraints will come from surveys that access Ly$\alpha$ emissions at high-$z$, such as HETDEX~\cite{Hill:2008mv} ($z_{\rm {Ly\alpha}} \sim 2$\,-\,$4$) and SPHEREx~\cite{SPHEREx:2014bgr} ($z_{\rm {Ly\alpha}}\sim 5$\,-\,$10$). Low redshift DDM decays contribute as interlopers to the flux observed by LIM surveys; at the same time, their radiation can be captured by our CBR analysis in the UV-band. Thence, the two techniques offer the opportunity to analyse DDM signatures with independent and uncorrelated probes.

\begin{figure}[ht!]
 \includegraphics[width=\columnwidth]{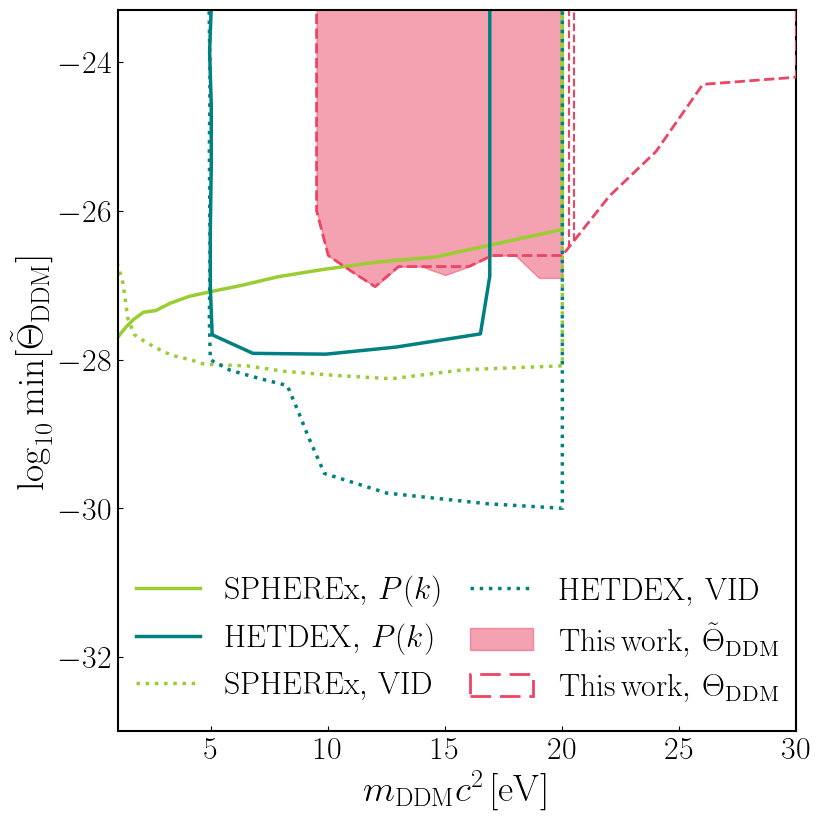}
    \caption{
    Minimum $\tilde{\Theta}_{\rm DDM} = f_{\rm esc}^{\rm DDM}f_{\gamma\gamma}f_{\rm DDM}\tau_{\rm DDM}^{-1}$ that can be detected at 95\% confidence level using the CBR technique on (GALEX+ULTRASAT)$\times$DESI (solid bar). Our results are compared with forecasts for LIM surveys from Ref.~\cite{Bernal:2020lkd}, using the one- (dotted) and two- (solid) point correlation function. The dashed line shows our results for (GALEX+ULTRASAT)$\times$DESI on ${\Theta}_{\rm DDM} = f_{\gamma\gamma}f_{\rm DDM}\tau_{\rm DDM}^{-1}$, where $f_{\rm esc}^{\rm DDM}$, $e^{-\zeta(\nu,z)}$ are included in the model for the DDM emissivity, as described in Sec.~\ref{sec:EBL}, and constrained in Sec.~\ref{sec:analysis}. The dashed vertical lines indicate decays in the proximity of the Ly$\alpha$ line, as in Fig.~\ref{fig:signal_evolution}.} 
    \label{fig:theta_DDM_compare}
\end{figure}

The results we discussed so far, however, cannot be directly compared to the analysis in \hyperlink{B20}{\color{magenta} B20}; in fact, in our model in Sec.~\ref{sec:EBL}, the escape fraction $f_{\rm esc}^{\rm DDM}(\nu,z)$ and the absorption factor $e^{-\zeta(\nu,z)}$ lower the emissivity of the DDM decays; in \hyperlink{B20}{\color{magenta} B20}, instead, these quantities are included in the collective parameter $\Theta_{\rm DDM}$. 
In order to compare the constraining power of our CBR analysis with their results, we run again the Fisher matrix, this time fixing both $f_{\rm esc}^{\rm DDM}(\nu,z)\equiv 1,\,e^{-\zeta(\nu,z)}\equiv 1$;  this is analogous to including uncertainties on these quantities inside a {new parameter}, $\tilde{\Theta}_{\rm DDM}$, analogous to the one in \hyperlink{B20}{\color{magenta} B20}. Moreover, as in \hyperlink{B20}{\color{magenta} B20}, we stop our analysis at $m_{\rm DDM}c^2 < 20\,{\rm eV}$, assuming that all photons emitted in the DDM$\to \gamma+\gamma$ process by particles above this mass get absorbed before reaching the observer. For each of the remaining DDM masses that (GALEX+ULTRASAT)$\times$DESI can observe, we estimate the minimum $\tilde{\Theta}_{\rm DDM}$ detectable at $2\sigma$ as the smallest value of $\tilde{\Theta}_{\rm DDM}$ for which the Fisher forecast returns $\sigma_{\tilde{\Theta}_{\rm DDM}} \leq \tilde{\Theta}_{\rm DDM}/2$. 

Results are shown in Fig.~\ref{fig:theta_DDM_compare} and compared to forecasts for upcoming LIM surveys; we include the forecasts on $\Theta_{\rm DDM}$ from the previous section, for completeness. On the small mass end, the two approaches are equivalent, due to the shape of $f^{\rm DDM}_{\rm esc}$ (see Fig.~\ref{fig:DM_esc}).
Our forecast CBR analysis on (GALEX+ULTRASAT)$\times$DESI\footnote{Results for ULTRASAT$\times$DESI can be obtained analogously.} provides competitive and independent results with respect to upcoming LIM surveys. The modeling we introduced in Sec.~\ref{sec:EBL} allows us to push our results to higher masses than the prescription adopted in previous works, at the cost of a lower constraining power, particularly at the higher masses. Even when $f_{\rm esc}^{\rm DDM}(\nu,z)$ and $e^{-\zeta(\nu,z)}$ are fixed to 1, the constraining power shows a similar mass-dependence to the one we previously described; this is due to the fact that the $f_{\rm LyC}$ factor in Eq.~\eqref{eq:fesc} still affects the astrophysical EBL, changing as a function of redshift the capability of CBR of detecting the decays.

%%%%%%%%%%%%%%%%%%%%%%%%%%%%%%%%%%
%%%%%%%%%%%%%%%%%%%%%%%%%%%%%%%%%%
%%%%%%%%%%%%%%%%%%%%%%%%%%%%%%%%%%
%%%%%%%%%%%%%%%%%%%%%%%%%%%%%%%%%%

\section{Constraining  axion-like particles}\label{sec:ALP}

Axion-like particles (ALP) are among the most interesting candidates for DDM, see e.g.,~Refs.~\cite{Preskill:1982cy,Berezhiani:1989fu,Abbott:1982af,Dine:1982ah,Cadamuro:2010cz,Cadamuro:2011fd} and references within for review. The production of such particles is related with spontaneous symmetry breaking in the early Universe or in the hot stellar interiors. The mass of the ALP is set by the scale at which the symmetry breaks; if their mass is above $\sim 10^{-2}\,{\rm eV}$, these particles were in thermal equilibrium at early times and could constitute today a non-negligible part of the DM~\cite{Turner:1986tb}.
Interactions between ALPs and particles in the standard model are strongly suppressed, but the possibility exists that ALPs decay into two photons; the strength of the coupling, labelled as $g_{\rm DDM,\gamma\gamma}$, sets the decay lifetime 
\begin{equation}\label{eq:tau_alp}
    \tau_{\rm DDM} = \frac{16(4\pi \hbar)}{(m_{\rm DDM}c^2)^3 g_{\rm DDM,\gamma\gamma}^2}.
\end{equation}

Constraints on ALPs come from a bunch of cosmological and astrophysical observables; we collect in Fig.~\ref{fig:final_forecast} some of the most relevant probes that can be useful in the $10^{-1}$\,-\,$10^6\,{\rm eV}$ range. Blue regions in the figure are set by cosmological probes, and they are based on the assumption that ALPs constitute the DM. Such limits come,~e.g.~from DM abundance~(Ref.~\cite{Cadamuro:2011fd} constrains the ALP density based on the DM density measured by WMAP~\cite{WMAP:2010qai}), CMB spectral distortions~\cite{Capozzi:2023xie,Liu:2023nct}, light element production during the Big-Bang nucleosynthesis~\cite{Depta:2020wmr}, limit on the DM freeze-in mechanism~\cite{Langhoff:2022bij}. The lack of observations of certain line emissions in the DM halo surrounding our Galaxy (JWST,~\cite{Janish:2023kvi,Bessho:2022yyu,
Yin:2024lla}) or in galaxy clusters (VIMOS,~\cite{Grin:2006aw}), as well as the presence of neutral gas that survives heat injection inside dwarf galaxies~\cite{Todarello:2023hdk,Wadekar:2021qae}, place further constraints on the ALP scenario. On the other hand, green bounds in the figure are related with the possibility that ALPs are produced inside stellar interiors. These can be understood thinking that the increased rate of energy loss these particles would lead, may change both the internal stellar structure and processes that drive stellar evolution. As a consequence, this may change the ratio between different stellar populations in globular clusters~\cite{Ayala:2014pea,Dolan:2022kul}, as well as the relation between the stellar initial mass function and the expected mass of white dwarfs~\cite{Dolan:2021rya}; moreover, ALP constraints can come from Sun observations~\cite{DeRocco:2022jyq}.
Furthermore, limits are set by direct detection of solar neutrinos~\cite{Vinyoles:2015aba}, neutrinos and gamma-rays from the 1987A supernova~\cite{Balazs:2022tjl,Muller:2023vjm,Hoof:2022xbe,Diamond:2023scc}, or from the properties of low energy supernova explosion~\cite{Caputo:2022mah}. 
Finally, constraints are set by analysing the possibility that DM decays contribute to the EBL in different bandwidths~\cite{Cadamuro:2011fd,Bernal:2022wsu,Fong:2024qeq}, which can also cause attenuation of the $\gamma$-ray background~\cite{Bernal:2022xyi} and introduce further anisotropies in the EBL angular power spectrum~\cite{Gong:2015hke,Kalashev:2018bra,Nakayama:2022jza,Carenza:2023qxh}. 

By looking at Fig.~\ref{fig:final_forecast}, it is evident that the $\mathcal{O}(10\,{\rm eV})$ mass range is yet poorly constrained. This, as we described in Sec.~\ref{sec:EBL}, is due to difficulties in the detection of UV photons produced in the decay of ALPs in this mass range.
In this context, the forecasted constraints we obtained in Sec.~\ref{sec:analysis} using the CBR technique on a UV-broadband survey can provide a strong improvement, thanks to its sensitivity to the integrated emission from a broad redshift range, and the capability of reconstructing the redshift evolution of its intensity. 

We can therefore re-express the DDM constraints that we obtained in the previous section in the ALP $(m_{\rm DDM}c^2,g_{\rm DDM,\gamma\gamma})$ plane. 
The results we obtained on $\tau_{\rm DDM}$ via $\Theta_{\rm DDM}$ (and $\tilde{\Theta}_{\rm DDM}$) can be translated by the means of Eq.~\eqref{eq:tau_alp}, once that values for $f_{\gamma\gamma}$, $f_{\rm DDM}$ are chosen. We show in Fig.~\ref{fig:final_forecast} the forecasted minimum value of $g_{\rm DDM,\gamma\gamma}$ that CBR will be able to measure or rule out when $\Theta_{\rm DDM}$ is measured from (GALEX+ULTRASAT)$\times$DESI diffuse light maps. We set $f_{\gamma\gamma}  = 1,\,f_{\rm DDM} = 1$; constraints on other DM fractions, as well as on $\tilde{\Theta}_{\rm DDM}$ and for ULTRASAT$\times$DESI, can be obtained scaling the aforementioned results.

%%%%%%%%%%%%%%%%%%%%%%%%%%%%%%%%%%
%%%%%%%%%%%%%%%%%%%%%%%%%%%%%%%%%%
%%%%%%%%%%%%%%%%%%%%%%%%%%%%%%%%%%
%%%%%%%%%%%%%%%%%%%%%%%%%%%%%%%%%%

\subsection{Extended ALP mass spectrum}

As a further test, we check the possibility that ALPs have an extended mass spectrum. This case has been investigated in Ref.~\cite{Gong:2015hke}, which in turn was motivated by high-energy physics theoretical and experimental works e.g.,~Refs.~\cite{Svrcek:2006yi,Arvanitaki:2009fg}, to explain the excess in the intensity power spectra of the optical and near-IR extragalactic background light measured by the Hubble Space Telescope (HST,~\cite{Mitchell-Wynne:2015rha}) and the Cosmic Infrared Background Experiment (CIBER,~\cite{Zemcov:2017dwy}). 

\bigskip
\begin{figure}[ht!]
    \centering
\includegraphics[width=\columnwidth]{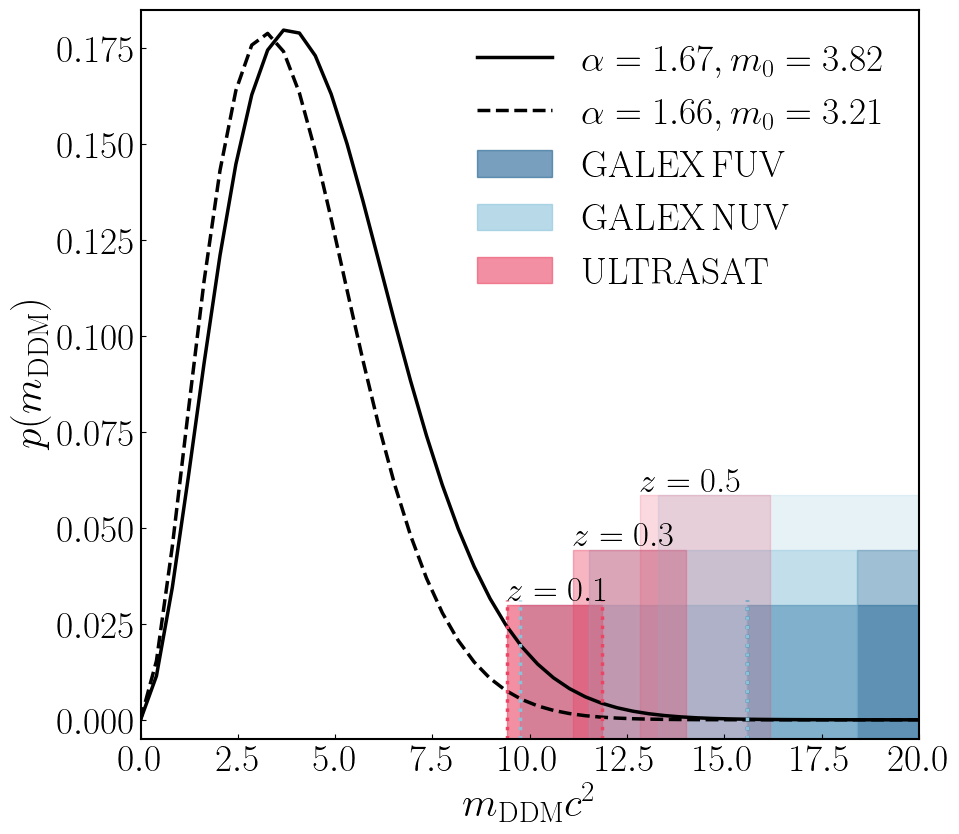}
    \caption{ALP mass spectrum in Eq.~\eqref{eq:p_mdm}. The parameters $\alpha,m_0$ are set to the best-fit values in Ref.~\cite{Gong:2015hke}; the two lines refer to different foreground models in their analysis. 
    Even if the mass spectrum determines mainly emission in the optical-IR band, we expect some contribution in the UV range: shaded areas indicate the masses ($\sim 9.5$\,-\,$17\,{\rm eV}$) of the particles that, at $z = \{0.1,0.3,0.5\}$, would be decaying into photons that can be observed by GALEX and ULTRASAT. Larger masses and higher redshifts have negligible contributions.}
    \label{fig:mass_spectr}
\end{figure}

\begin{figure*}[ht!]
    \centering
    \includegraphics[width=1.5\columnwidth]{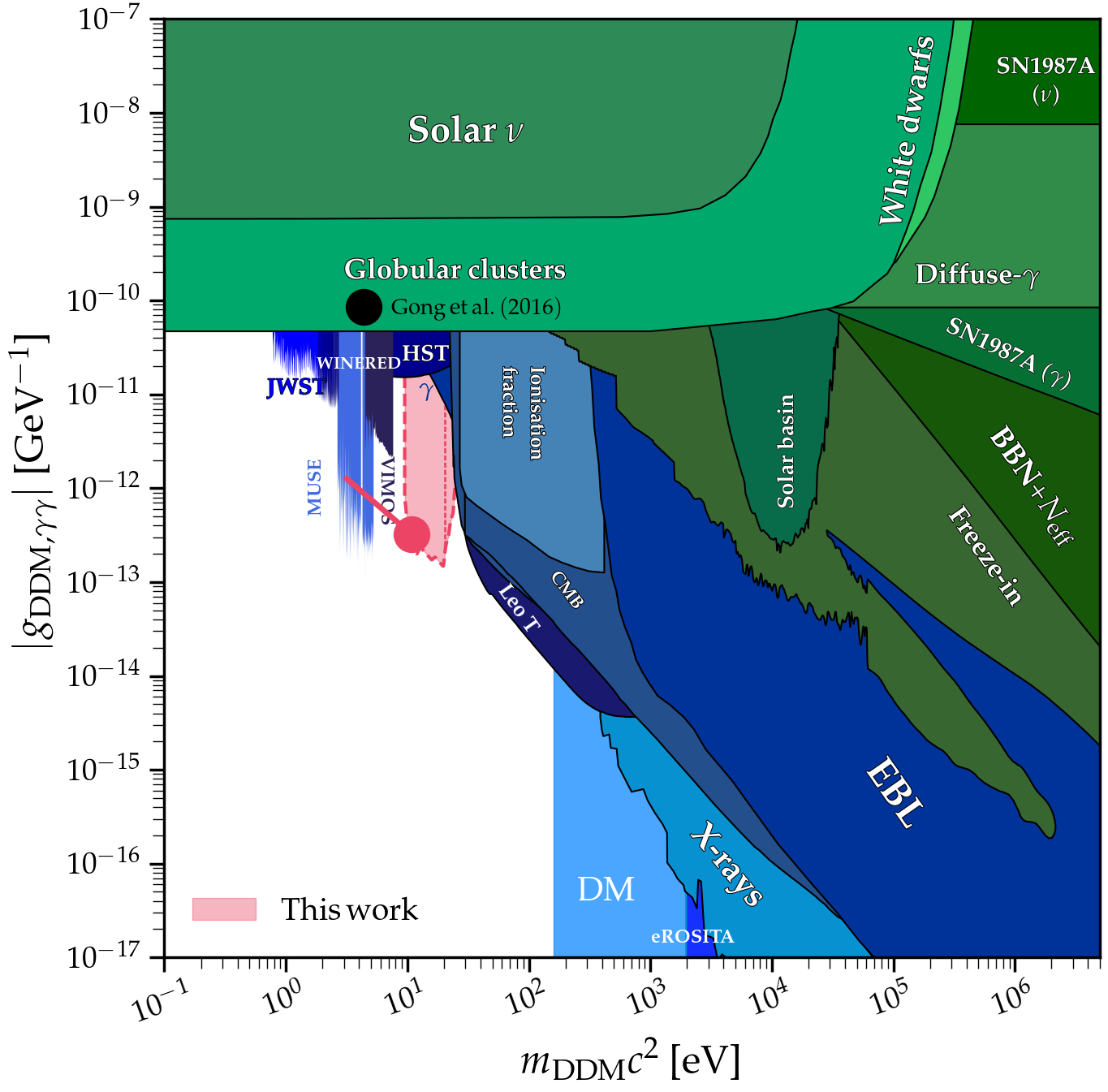}
    \caption{1$\sigma$ constraints on ALP mass and coupling constant. Forecasts from our work are indicated by the pink area, and based on constraints on $\Theta_{\rm DDM}$ obtained in Sec.~\ref{sec:analysis}, assuming $f_{\rm DDM} = 1,\,f_{\gamma\gamma}=1$, and converted to $g_{\rm DDM,\gamma\gamma}$ through Eq.~\eqref{eq:tau_alp}. 
    Blue regions assume that ALPs constitute the DM, while green areas are associated with the possibility that ALPs are produced inside stellar interiors. Constraints are described in the main text and plotted using libraries in \url{https://github.com/cajohare/AxionLimits}. The black dot indicates the constrain Ref.~\cite{Gong:2015hke} obtained from HST and CIBER power spectrum amplitudes using the ALP mass spectrum in Eq.~\eqref{eq:p_mdm}. If that was correct, the tail of the distribution would produce signatures also in our  CBR analysis of the UV-EBL. The constraining power from (GALEX+ULTRASAT)$\times$DESI, in this case, is shown by the pink dot and projected to lower masses through the pink line, assuming that all particles in the spectrum have the same decay time. Details in Sec.~\ref{sec:ALP}.  }
    \label{fig:final_forecast}
\end{figure*}

We follow the same formalism to model the ALP mass distribution function in a phenomenological way, as 
\begin{equation}\label{eq:p_mdm}
\begin{aligned}
    p(m_{\rm DDM}) =& \frac{\alpha}{m_0}\frac{1}{\Gamma(1+1/\alpha)}\left(\frac{m_{\rm DDM}}{m_0}\right)^\alpha\times\\
    &\times \exp\left[-\left(\frac{m_{\rm DDM}}{m_0}\right)^{\alpha}\right],
\end{aligned}
\end{equation}
where $\alpha,m_0$ are free parameters and $\Gamma(x)$ the Gamma function. 
Ref.~\cite{Gong:2015hke} estimated $\alpha=1.67\,(1.66),\,m_0=3.82\,(3.21)$ as best fit values (using different assumptions on the foregrounds). We show with a black dot on Fig.~\ref{fig:final_forecast} the results obtained in Ref.~\cite{Gong:2015hke} for the extended mass model, using the fiducial values $\alpha=1.67,m_0=3.82$. The position along the $x$-axis indicates the mass where the distribution peaks.

If ALPs are indeed described by the mass function in Eq.~\eqref{eq:p_mdm}, we expect to detect signatures in the UV-EBL emissivity analysed in this work. The broadband frequency range observed by ULTRASAT and GALEX will capture emissions from different $z$, sourced by the high-mass tail of the distribution, as we show in Fig.~\ref{fig:mass_spectr}. 
To put constraints on this possibility, we extend our CBR forecast to the ALP extended mass spectrum scenario: to do so, we account for $p(m_{\rm DDM})$ in the computation of the DDM emissivity in Eq.~\eqref{eq:eps_DM}. Since the broadband surveys under analysis integrate the observed signal in $\nu_{\rm obs}$, the contribution from the extended ALP mass spectrum simply combine in the final result. Different masses will contribute to the UV-EBL from different redshifts, with relative weights estimated from the mass distribution. 

We update the computation of ${d{J}_{\nu_{\rm obs}}}/{dz}$ in Eq.~\eqref{eq:dJnu} accounting for the extended mass spectrum and we run the Fisher matrix from Eq.~\eqref{eq:fisher}, similarly to the previous section. In this way, we estimate the minimum $\Theta_{\rm DDM}$ that can be detected at $1\sigma$.\footnote{We could analogously run the analysis on $\tilde{\Theta}_{\rm DDM}$, changing the assumptions on the escape fraction.} This is then converted to $g_{\rm DDM,\gamma\gamma}$ by setting $f_{\rm DDM} = 1, f_{\gamma\gamma}=1$ and inverting Eq.~\eqref{eq:tau_alp}. The forecasted value of $g_{\rm DDM,\gamma\gamma}$ that can be detected at 1$\sigma$ is shown in Fig.~\ref{fig:final_forecast} using a pink dot; we associate it to the average mass observed in the GALEX+ULTRASAT bandwidth, which we estimate as 
\begin{equation}
   m_{\rm DDM}|_{p(m_{\rm DDM})} = \frac{\int_{9.5\,{\rm eV}}^{30\,{\rm eV}}{dm_{\rm DDM}\,p(m_{\rm DDM})m_{\rm DDM}}} {\int_{9.5\,{\rm eV}}^{30\,{\rm eV}}{dm_{\rm DDM}\,p(m_{\rm DDM})}}.
\end{equation}

The constraints CBR will provide on the ALP high mass tail can be extended to the full spectrum, under the assumption that all the particles have the same decay rate. Therefore, we use Eq.~\eqref{eq:tau_alp} to project our results to the mass range probed by Ref.~\cite{Gong:2015hke}; we compare our forecasts with their results through the pink line in the figure. Our forecasts show that CBR on the UV-EBL will be able to detect or rule out emissions from the extended ALP mass spectrum proposed in Ref.~\cite{Gong:2015hke}.

\section{Conclusions}
\vspace*{-.3cm}

Cosmology nowadays relies on the $\Lambda$CDM model, which is capable of explaining most of the current observations. Many questions, however, are still unanswered; not least, what is the nature of the dark matter component that determines the gravitational potential on large scales and drives structure formation. 

Among the different classes of possible explanations, many account for the chance that DM is made up of  beyond  Standard Model particles~\cite{Bertone:2004pz,Profumo:2017hqp,Schumann:2019eaa,Gaskins:2016cha}, capable of weakly interacting with regular particles, to some extent. If this is the case, at least part of the DM particles may be able to decay into photons, thus providing energy injections that can be detected with cosmological observables. 

One of the best ways we have to detect decaying DM components, therefore, is to refer to instruments capable of collecting the integrated emissions from a wide field of view, without the need of resolving each single source. 
For this reason, many of the current constraints on DDM are based on measurements of the extragalactic background light (EBL), and on the study of its anisotropies~\cite{Cadamuro:2011fd,Gong:2015hke,Kalashev:2018bra,Nakayama:2022jza,Carenza:2023qxh}. 
It has been shown, see e.g.,~Refs.~\cite{Bernal:2022xyi,Bernal:2022wsu,Bernal:2022xyi,Creque-Sarbinowski:2018ebl,Liu:2019bbm,Facchinetti:2023slb}, that upcoming line-intensity mapping (LIM) surveys also have huge potential in this direction. 

The aim of this work has been to explore the possibility that broadband UV detectors, such as GALEX~\cite{Martin:2004yr,Morrissey:2007hv} or ULTRASAT~\cite{Sagiv:2013rma,ULTRASAT:2022,Shvartzvald:2023ofi}, may be capable of placing new and independent constraints on the properties of DDM. We focused on the use of cross correlation with spectroscopic galaxy surveys,~e.g.,~DESI~\cite{DESI:2013agm,DESI:2016fyo,DESI:2016igz,DESI:2023dwi}, as a tool to reconstruct the redshift evolution of the UV intensity and comoving emissivity, using the so-called clustering-based redshift (CBR,~\cite{Newman:2008mb,McQuinn:2013ib,Menard:2013aaa}) method.
The use of CBR allows us to overcome the difficulties that observations in the UV band present, due to absorptions in the intergalactic medium (IGM) and to the presence of foregrounds. In this work, we assumed to rely on the full-sky map that ULTRASAT will build in the first six months of the mission; a possible alternative is to use data from the low-cadence survey realized along the full duration of the mission. This will scan a smaller sky area, reaching $\sim 10$\,-\,$100$ times better sensitivity; Ref.~\cite{Libanore:2024wvv} showed that the maps produced in these two cases will provide a similar constraining power in the context of CBR analysis.

In our work, we considered DDM$\to \gamma+\gamma$ decays, in which a DDM particle decays to two monoenergetic photons, whose frequency is set by the mass of DDM.
In agreement with the observational bandwidth of GALEX and ULTRASAT, and according to the redshift range probed by DESI, our analysis is capable of constraining particles with masses between 9.5\,eV and 30\,eV; while the lower bound is set by the maximum wavelength in the observational range, the upper bound has been computed accounting for efficient absorptions when photons have energy close to the hydrogen ionization level. Particles that decay into photons with energy close to the Ly$\alpha$ line should be excluded from the analysis, since the intensity of the line may overcome the decay contributions. 

We modeled the comoving-volume emissivity of DDM following Ref.~\cite{Bernal:2020lkd} (\hyperlink{B20}{\color{magenta} B20} in the paper), and we combined it with the UV-EBL emissivity with astrophysical origin, modeled in Ref.~\cite{Chiang:2018miw}\,(\hyperlink{C19}{\color{magenta} C19}). We followed the procedure detailed in Ref.~\cite{Libanore:2024wvv} (\hyperlink{LK24}{\color{magenta} LK24}), to perform a Fisher forecast analysis on the CBR-reconstructed, redshift-dependent UV intensity and to estimate the minimum DDM decay rate that (GALEX+ULTRASAT)$\times$DESI will be able to constrain at 2$\sigma$. Our results obviously depend on the particle mass, and encapsulate the degeneracy between the decay rate, the branching ratio of the decaying process, and the DDM abundance, expressed in terms of the DM fraction made by this kind of particles. 
To perform our Fisher forecast, we assumed that the GALEX NUV$\times$DESI, GALEX FUX$\times$DESI and ULTRASAT$\times$DESI datasets are uncorrelated, while analyses on real maps instead will require the computation of the covariance between the experiments, which may worsen the final results. For this reason, we also quoted results for ULTRASAT$\times$DESI alone, which are almost an order of magnitude worse than the combined ones. Actual results will reasonably fall between our two forecasts. 
A more refined computation of the noise term in Sec.~\ref{sec:noise} is also needed; however, since the specs we adopted for ULTRASAT follow conservative choices, observational improvements may also compensate for this fact (e.g.,~we only considered the first six months of the mission to estimate the full-sky map sensitivity, while follow-up data can also be added; moreover, we neglected the use of tithering and overlapping patches in the CBR).

Fig.~\ref{fig:theta_DDM_compare} summarizes our forecast results, and compares them with the constraints from upcoming LIM surveys obtained in Ref.~\cite{Bernal:2020lkd}, in a similar DDM mass range. The CBR technique will provide competitive and independent constraints with respect to LIM surveys, establishing itself as a complementary tool in the study of the EBL, from both astrophysical and cosmological origins. 

Finally, we studied the case in which DDM is made of  axion-like particles (ALPs,~\cite{Preskill:1982cy,Abbott:1982af,Dine:1982ah,Cadamuro:2010cz,Cadamuro:2011fd}), and we converted our results into forecasts on the ALP-photon coupling constant accessible by broadband UV surveys, when analysed using CBR. We considered both the cases in which ALPs are described by a monochromatic mass function, and by the broad mass distribution in Ref.~\cite{Gong:2015hke}.  

In Fig.~\ref{fig:final_forecast} we show our forecast results compared with a summary of current constraints on ALPs. Interestingly, our technique will probe a mass window which is still largely unconstrained, being also able to detect or rule out emissions due to the ALP mass spectrum in Ref.~\cite{Gong:2015hke}. The improvement in our analysis comes from the fact that, being able to reconstruct the redshift evolution of the UV-EBL intensity, the cross correlation between GALEX, ULTRASAT and DESI makes the analysis more sensitive to ALPs (or, in general, to DDM) contributions than just the quest for excess in the UV signal with respect to the only-astrophysical contribution. 

To conclude, we highlight the importance of using independent probes and regimes to constrain the properties of DDM. The use of UV detectors, combined with high-$z$ LIM, represents a great opportunity in this sense.

\vspace*{-.5cm}
\acknowledgments
\vspace*{-.3cm}
\noindent The authors are extremely thankful to José L. Bernal for his insightful reading and for his thorough suggestions on the analysis and text. 
SL is supported by an Azrieli International Postdoctoral Fellowship. EDK was supported by a Faculty Fellowship from the Azrieli Foundation. EDK also acknowledges joint support from the U.S.-Israel Bi-national Science Foundation (BSF,  grant No. 2022743) and the U.S.\ National Science Foundation (NSF, grant No. 2307354), and support from the ISF-NSFC joint research program (grant No. 3156/23). 
We acknoweldge the use of libraries in~\url{https://github.com/cajohare/AxionLimits} to produce Fig.~\ref{fig:final_forecast}.

\end{document}